\documentclass{article}

\usepackage{PRIMEarxiv}

\usepackage[utf8]{inputenc} 
\usepackage[T1]{fontenc}    
\usepackage{booktabs}       
\usepackage{amsfonts}       
\usepackage{nicefrac}       
\usepackage{microtype}      
\usepackage{fancyhdr}       
\usepackage{graphicx}       
\graphicspath{{figures/}}     
\usepackage{subcaption}
\usepackage{algpseudocode}
\usepackage{cite}
\usepackage[linesnumbered,ruled,norelsize]{algorithm2e}
\usepackage{amsmath,amssymb,amsfonts}

\title{A Reinforcement Learning Approach to Optimize Available Network Bandwidth Utilization\\
}

\author{
  Hasibul Jamil, Elvis Rodrigues, Jacob Goldverg, Tevfik Kosar\\
  Department of Computer Science and Engineering \\
  University at Buffalo (SUNY)\\
  Amherst, NY 14260\\  
  \texttt{\{mdhasibu, elvisdav, jacobgol, tkosar\}@buffalo.edu} \\
}

\begin{document}
\maketitle

\begin{abstract}
Efficient data transfers over high-speed, long-distance shared networks require proper utilization of available network bandwidth. Using parallel TCP streams enables an application to utilize network parallelism and can improve transfer throughput; however, finding the optimum number of parallel TCP streams is challenging due to nondeterministic background traffic sharing the same network. Additionally, the non-stationary, multi-objectiveness, and partially-observable nature of network signals in the host systems add extra complexity in finding the current network condition. In this work, we present a novel approach to finding the optimum number of parallel TCP streams using deep reinforcement learning (RL). We devise a learning-based algorithm capable of generalizing different network conditions and utilizing the available network bandwidth intelligently. Contrary to rule-based heuristics that do not generalize well in unknown network scenarios, our RL-based solution can dynamically discover and adapt the parallel TCP stream numbers to maximize the network bandwidth utilization without congesting the network and ensure fairness among contending transfers. We extensively evaluated our RL-based algorithm's performance, comparing it with several state-of-the-art online optimization algorithms. The results show that our RL-based algorithm can find near-optimal solutions 40\% faster while achieving up to 15\% higher throughput. We also show that, unlike a greedy algorithm, our devised RL-based algorithm can avoid network congestion and fairly share the available network resources among contending transfers.
\end{abstract}


\section{Introduction}
The exponential growth of scientific and commercial data and the need to infer knowledge from it often require moving large amounts of data between geographically separated institutions. For example, the National Energy Research Scientific Computing Center (NERSC) generates close to 1 Exabyte of data per year, which needs to be transferred to Oak Ridge Leadership Computing Facility for petascale simulations~\cite{LADS2015}. As a result, high-performance computing and scientific communities strive for large amounts of network bandwidth. But, even after being given high-speed networks, most data transfers can only reach a few Gbps because of improper network bandwidth utilization. It has been shown ~\cite{jamil-ICCCN-2022,prasanna-2016,altman-2006,Dong-2005,hacker2002, kim2015highly} that single stream TCP throughput achieved by data transfer applications is a fraction of the available end-to-end network bandwidth, and this deficiency is due to TCP's additive increase and multiplicative decrease (AIMD) window control procedure. 

There are mainly two types of congestion control mechanisms: (1) loss-based algorithms \cite{cubic2008,newReno2004} can obtain higher throughput, but they suffer from higher round trip time (RTT); (2) delay-based algorithms \cite{copa2018,FastTCP} are usually subject to acknowledgment (ACK) delay and network jitter, and these algorithms often result in network capacity under-utilization. The throughput deficiency of a TCP stream is subject to a window control mechanism in TCP congestion avoidance in response to packet loss. To illustrate the effect of packet loss on TCP throughput,  TCP bandwidth estimation equations by Mathis~\cite{mathis1997} could be used where it is shown that TCP throughput is inversely proportional to the square root of the packet loss rate. Because of that, a minimal packet loss rate is required for a single TCP stream to achieve close to full network capacity usage on high Bandwidth Delay Product (BDP) networks. For example, the packet loss rate needs to be less than or equal to  0.0018\% to allow a single TCP stream to utilize at least 2/3 of a 622 Mbps Optical Carrier (OC-12) ATM link~\cite{Othman-2013}. It is important to note that achieving such a low packet loss rate is not very practical as packet loss could be due to network congestion, congestion control procedure, or random event from different hardware failures~\cite{Lakshman-1997}.

A common approach to improve the throughput of data transfer is to discover a mechanism where applications utilize a set of parallel TCP streams to increase the aggregated TCP throughput and achieve higher network bandwidth utilization. Applications with multiple parallel TCP streams could better use available network resources by rapidly scaling up to peak bandwidth and responding to random packet loss less aggressively~\cite{prasanna-2016}. On the other hand, multiple TCP streams could incur computational and memory overhead in the end system, competing unfairly with different flows that share the same network link/links. So too little use of parallel TCP streams will underutilize the available network bandwidth, but too much use will overburden the end systems and cause congestion. Therefore, this problem is multi-objective; not only do we need to utilize the available network capacity, but we also should avoid creating congestion in the network. 

\begin{figure*}[ht]
  \begin{subfigure}[b]{0.5\textwidth}
    \includegraphics[width=\columnwidth]{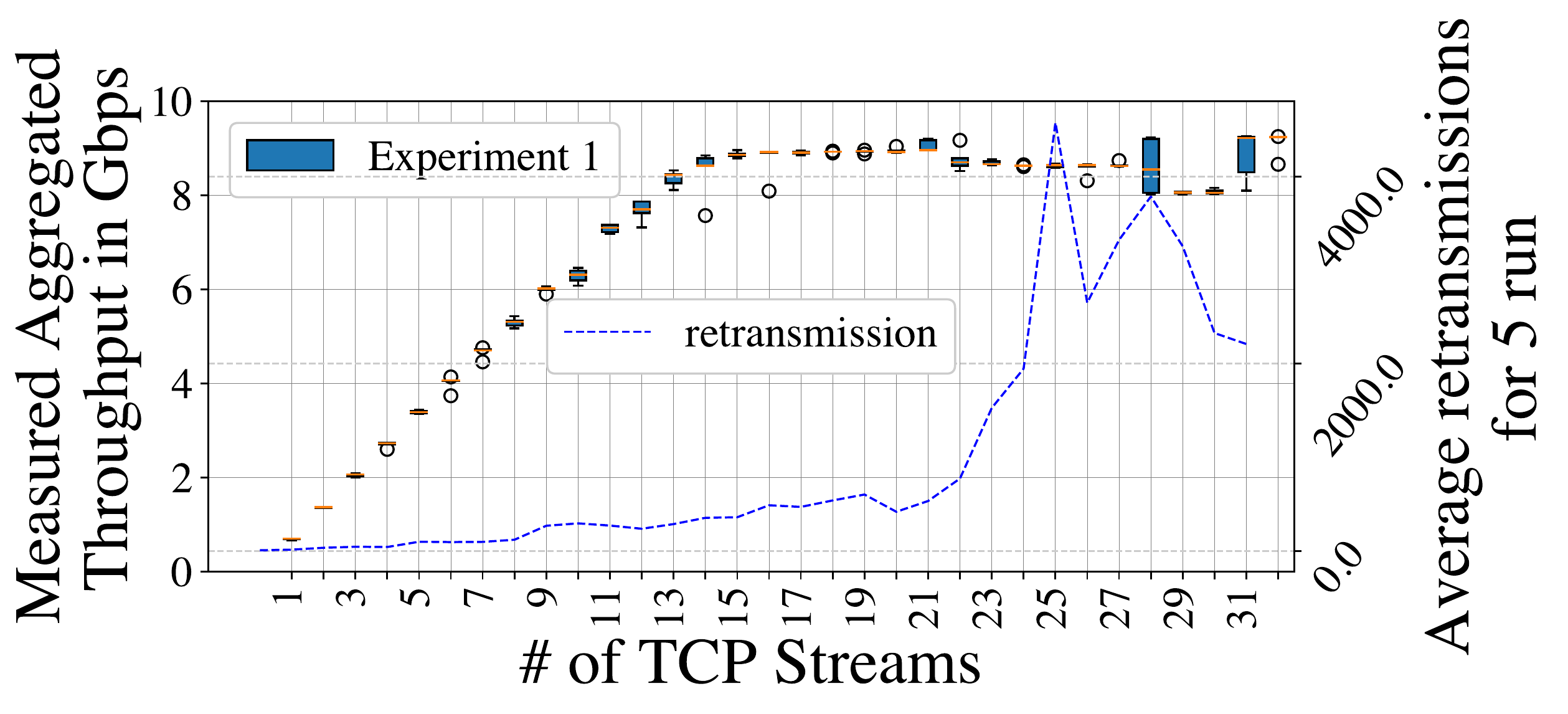}

    \caption{Experiment 1 between\\
     UC to Cloudlab Utah}
    \label{fig:tcp-throughput}
  \end{subfigure}
  \hspace{-0.20cm}
  \begin{subfigure}[b]{0.5\textwidth}
    \includegraphics[width=\columnwidth]{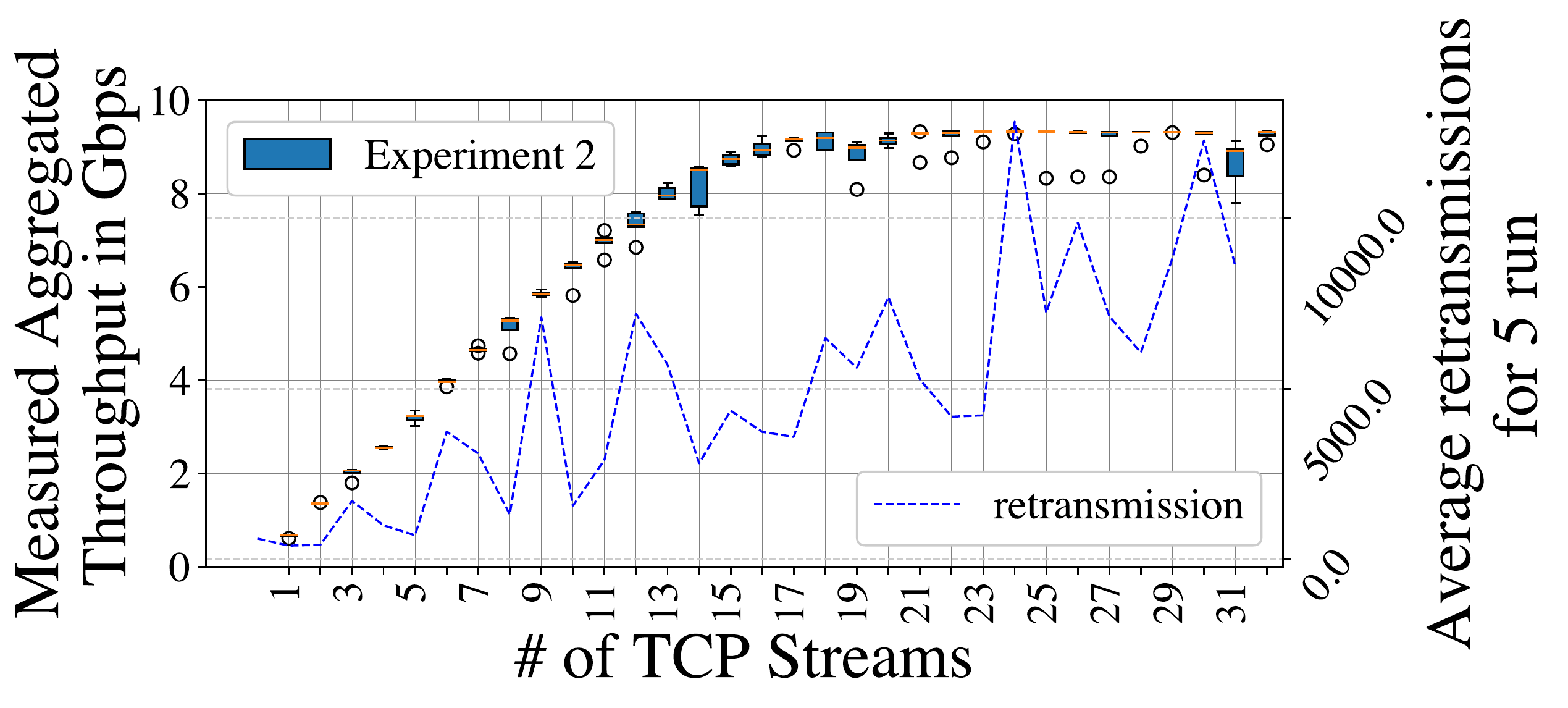}
    \caption{Experiment 2 between\\
     UC to Cloudlab Utah }
    \label{fig:tcp-throughput_2}
  \end{subfigure}
     \hfill 
  \begin{subfigure}[b]{0.5\textwidth}
    \includegraphics[width=\columnwidth]{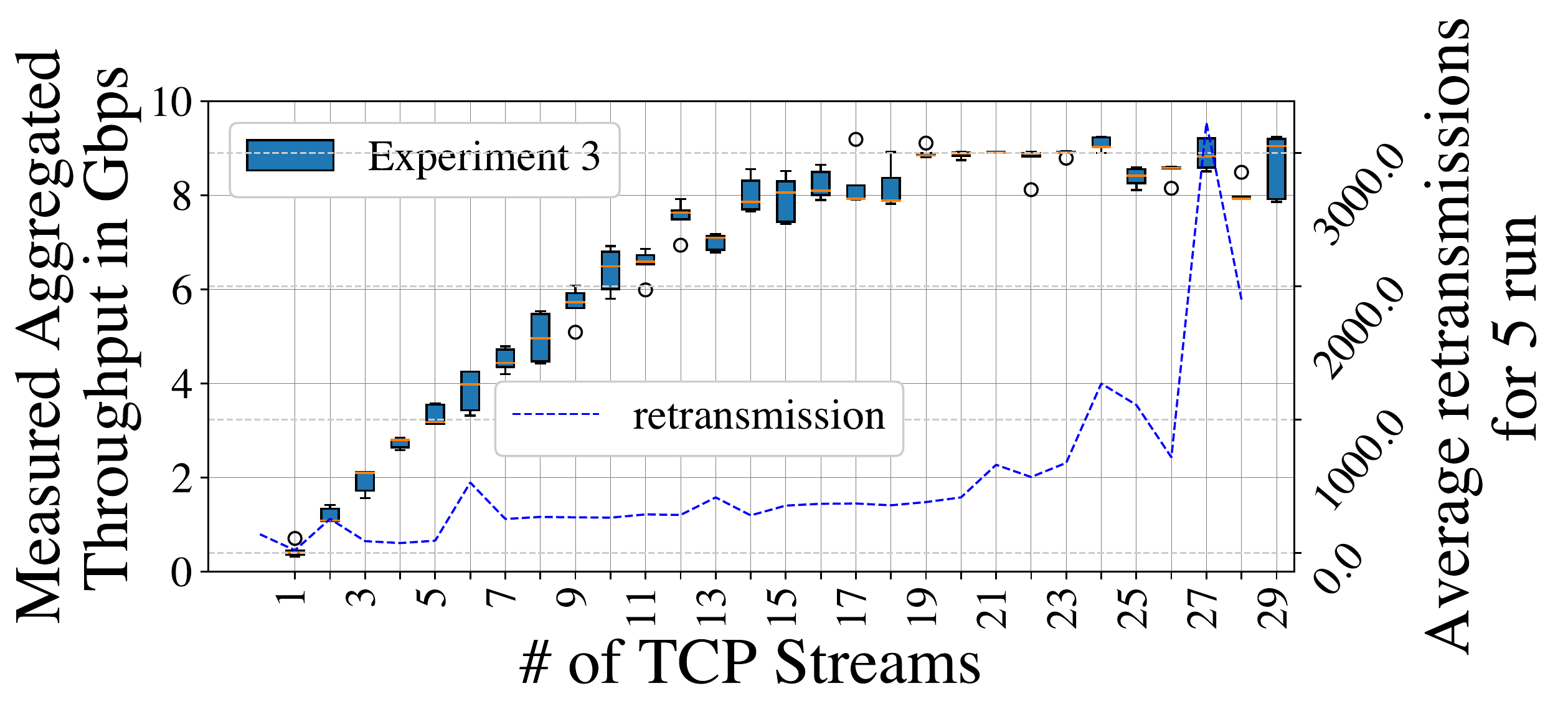}
    \caption{Experiment 3 between\\
     UC to TACC }
    \label{fig:tcp-throughput_3}
  \end{subfigure}
  \hspace{-0.20cm}
  \begin{subfigure}[b]{0.5\textwidth}
    \includegraphics[width=\columnwidth]{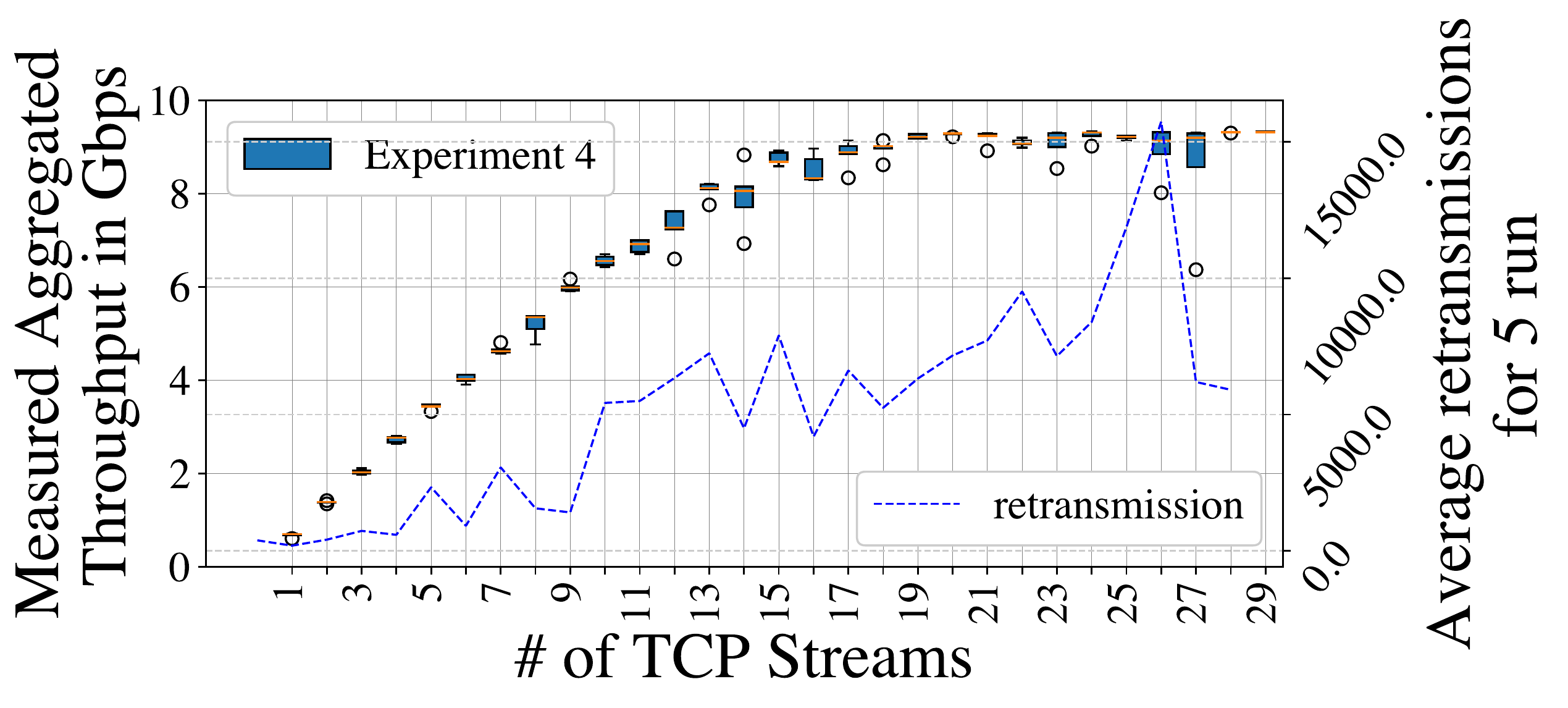}
    \caption{Experiment 4 between\\
     UC to TACC }
    \label{fig:tcp-throughput_4}
  \end{subfigure}
  \caption{Aggregated Throughput and average re-transmission number with multiple parallel TCP (CUBIC) streams. Each data point of stream value has been tested 5 times and the plot shows the statistical distribution of the aggregated throughput results and the average number of re-transmission. }
  \label{fig:figure1}
\end{figure*}

To demonstrate the impact of parallel TCP streams on achieved throughput and network congestion, we tested a file transfer application with a varying number of parallel TCP streams on a 10 Gbps link between Chameleon Cloud~\cite{keahey2020lessons} UC node and Cloudlab~\cite{Duplyakin+:ATC19} Utah node. Figure~\ref{fig:figure1} shows that maximum aggregated transfer throughput is achieved when 15 parallel TCP streams are used (Figure ~\ref{fig:tcp-throughput}). Beyond that point, adding additional TCP streams will not improve the aggregate throughput performance but instead, hurt the performance and dramatically increase the re-transmission due to packet loss from a congested network. Figure ~\ref{fig:tcp-throughput_2} shows that for the same endpoints, transfer conducted in a different time and duration, the optimum number of parallel TCP streams is 18. Similar behavior is observed in other network endpoints, i.e., between Chameleon UC and Chameleon TACC nodes, as shown in Figure ~\ref{fig:tcp-throughput_3} and \ref{fig:tcp-throughput_4} where the optimum number of parallel TCP stream is 20 and 19 respectively. As we can see, the optimal number of parallel TCP streams is not fixed for all transfers as it depends on different dynamic conditions (e.g., network background traffic, network route, endpoints, and network configuration). These particular characteristics introduce the non-stationary part of the problem.
The transfer application is isolated from other processes on the same host and from other hosts currently sharing the network for security reasons, making the process only partially observable. To overcome these challenges, we formulate finding the optimal number of TCP streams as a sequential decision problem where the observation for decision-making is not known beforehand and model this problem as a partially observable Markov Decision Process (MDP). 

Previous work uses real-time probing ~\cite{Dong-2005}, heuristics ~\cite{di2019cross}, and historical analysis models ~\cite{jamil-ICCCN-2022,Rodolph2021} to find the optimum number of TCP streams. Heuristics models suffer from poor robustness as the network condition (e.g., background traffic and traffic pattern) they are designed for might not be present in other networks. Probing can significantly overburden the network resources with unnecessary traffic and potentially create unfairness among other contending background traffic. In this paper, we present a learning-based approach that utilizes deep reinforcement learning algorithms to devise a policy function approximator that can swiftly discover the optimum number of parallel TCP streams to maximize available network bandwidth utilization. The learned policy takes network signals available in the host node as the observation of the MDP state and takes action by adjusting the number of TCP streams to maximize the achieved throughput. Also, the proposed method ensures fairness among other contending transfers by incorporating a \textit{punishment} term in the utility function that defines the reward signal for our RL agent. The major contributions of this paper include the following:
\begin{itemize}
    \item To the best of our knowledge, we are first to show that aggregated TCP throughput maximization is a good fit for the RL problem and devise a policy gradient method to maximize available bandwidth utilization.
    \item We showed how our reward function guarantees to maximize throughput but at the same time ensures fairness among other contending background traffic.
    \item We extensively evaluated our solution with different online optimization algorithms and showed that our learning-based algorithm achieves, on average 40\% faster convergence and up to 15\% better-aggregated throughput. 
\end{itemize}
         
The rest of the paper is organized as follows: Section \textrm{II} presents the background and related works; Section \textrm{III} discusses our proposed method; Section \textrm{IV} presents the evaluation of our method; Section \textrm{V} concludes the paper.

\section{Background and Related Work}
In this section, starting from a single TCP stream model, we will see why multiple streams are required to maximize network bandwidth utilization. We will also shed light on how multiple streams could potentially congest the network and create fairness issues with other background traffic and how we can avoid that. We will also discuss several similar works and argue why our model is superior to other models. 

A single TCP stream throughput is modeled with the following equation~\ref{eq:singleTCPBW} in case the packet loss rate is less than 1/100 as shown by Mathis~\cite{mathis1997}:
\begin{equation}
\label{eq:singleTCPBW}
T C P_{t h r} \leq \frac{M S S}{R T T} \frac{C}{\sqrt{p}}
\end{equation}
Here, $TCP_{thr}$ is TCP throughput, $MSS$ is the maximum segment size, $RTT$ is the round trip time, $C$ is a constant and $p$ is the packet loss ratio which is the number of retransmitted packets divided by the total number of packets transmitted. For a TCP session, $MSS$ is static and depends on network structure, host operating system, host network card, and network switch characteristics. The value of $RTT$ has a lower bound of the speed of the signal from end hosts in the TCP session (i.e., client and server machine) and depends on queue conditions in the intermediary network devices and the congestion state of the network. The packet loss rate $p$ is the most important signal to the TCP congestion avoidance algorithm. If a packet loss is from network congestion or traffic insensitive reason (i.e., hardware or link failure, TCP random early detection), no matter for what reason packet loss happens, TCP congestion avoidance takes that packet loss as a signal for network congestion and subsequently decrease the TCP sending rate (i.e., decreasing TCP window size). We can extend the single stream TCP throughput model for multiple TCP streams by aggregating individual throughput of each stream to derive the application's aggregated $TCP_{agg}$ as shown in equation~\ref{eq:multiTCP1} \cite{hacker2002}:
\begin{equation}
\label{eq:multiTCP1}
T C P_{a g g} \leq \frac{C}{R T T}\left[\frac{M S S}{\sqrt{p 1}}+\cdots+\frac{M S S}{\sqrt{\mathrm{pn}}}\right]
\end{equation}
where $p_i$ represents the packet loss rate of TCP connection $i$, $MSS$ is the same as above and fixed for all TCP streams and $RTT$ is the converged round trip time value of all the TCP connections. As from equation ~\ref{eq:multiTCP1}, we can see that having multiple streams is equivalent to creating a $n$ times large virtual $MSS$ than one single stream, and because of this virtual jumbo $MSS$, multiple streams could achieve higher aggregated throughput given $p_i$ is close to each other for different values of $i$. This condition that the proportion of packet loss is evenly distributed across all streams is only true in the absence of congestion. 

However, when the network gets congested, whether by increasing the number of parallel TCP streams or the presence of other heavy background traffic, the benefits of adding additional TCP streams diminish. There are two reasons for that; first, the additional TCP stream will make the network congested, so the packet loss rate will increase, and this additional packet loss will offset any improvement over aggregate TCP throughput that could have been achieved by the additional TCP streams. Second, the sender and receiver connecting bottleneck links in the network are at full capacity so it can not offer any extra bandwidth. As a result, maximum available network bandwidth utilization will be possible if the application can operate on the sweet spot in terms of the number of TCP streams that do not put a network in a congestion state. As the background traffic in network infrastructure is dynamic, the optimal number of TCP streams is also not fixed and varies during the transfer session. Our DRL algorithm could learn a policy to intelligently avoid the congestion in the network and choose an optimum number of parallel TCP streams by utilizing different network signals available in end hosts to optimally use the available network bandwidth. 

There is a lot of work in proposing a new transport layer to maximize bandwidth utilization ~\cite{DataCenterCongestionControl-2022,qtcp,BBR2016,PCC-Vivace}. Still, wide adoption of transport layer protocol requires kernel and system modification, making this type of solution challenging. Dong et al.\cite{Dong-2005} propose probing techniques to estimate available bandwidth. Network probing introduces an additional burden to the production traffic, and too much probing in terms of duration and magnitude could create network congestion. Several ad-hoc heuristics methods~\cite{di2019cross} utilized observation-based rules to select the number of parallel streams but they suffer from a lack of robustness for new or unseen network scenarios. Falcon~\cite{arif2021}, Balaprakash et al~\cite{prasanna-2016} tackles the transfer throughput maximization problem as an online convex optimization which is stateless and it does not optimize for long term behavior rather focuses on immediate performance improvement. In our work, we present a learning-based approach that utilizes deep reinforcement learning algorithms to devise a policy function approximator that can generalize the relationship between observed network signals in the end host and network congestion state and discover the optimum number of parallel TCP streams to operate on the sweet spot to avoid congestion but maximally utilize available network bandwidth. 

\section{METHODOLOGY}
In this section, we explore the use of DRL to devise a policy function approximator that can learn the relationship between observed network signals in the end host and network congestion state and discover the optimum number of parallel TCP streams to maximally utilize available network bandwidth. Our learning-based solution could learn the mapping and model the dynamics between observed network signals in the end system and overall network congestion conditions from past experiences (e.g., during training). Overall our design principles are as follows (1) No prior knowledge about the network or end hosts is required; (2) Only observations obtainable in the end host are used to intelligently develop an intuitive understanding of the network condition. To achieve the design principles, we formulate the optimum network utilization problem as Markov Decision Process and devise a deep DRL solution that chooses a sequence of actions to maximize the cumulative payoff (i.e., return) of the individual chosen action. Each step of the sequence of actions could be considered as monitoring intervals(MIs)\cite{PCC-Vivace}. The environment consists of the end hosts, data transfer application, and the actual network links, while the DRL policy agent uses a Deep Neural Network (i.e., DNN) model that takes present and previous $n$ steps network signals as input state and selects the action either to increase, decrease or keep the number of parallel TCP streams same as the previous step. Environment executes the chosen action for a preset amount of time and gives back current and last $n$ previous performance metrics such as achieved throughput, packet loss rate ($plr$), and round trip time ($rtt$) as feedback. The performance metrics are used with a utility function to devise a reward for the chosen action of the policy agent, and over time the agent learns to optimize its decisions to maximize total reward. The feedback is then used to derive the input state for the RL for the next MIs (i.e., steps) and this cycle repeats until the transfer process finishes (i.e., end of the episode). 

\subsection{DRL Architecture}
\subsubsection{State Space}
In the RL paradigm, a larger set of attributes or state variables could make the size of the exploration space exponentially larger. It is hence desirable to minimize the number of attributes or state variables of the environment and only include those that closely relate to the DRL agent’s goal. We use the last $n$ including the current step of network signals. Because different network link characteristics such as loss rate, available bandwidth, and latency are different, the state variables in terms of network signal should avoid those signals that have high variability (e.g., choosing change in $RTT$ over time rather than the absolute value of $RTT$ in milliseconds). The above attribute of network signals as state variables ensures an enhancement in the generalization capability of the DRL model across different network scenarios. Sender selects action $a_t$ at MI $t$, after a preset time, it gets performance metrics as observation feedback for choosing that particular action and computes a signal vector $x_t$ from the feedback, consisting following state variables: (i) $RTT$ gradient, the derivative of $RTT$ with respect to time; (ii) $RTT$ ratio, the ratio of the current MI’s mean $RTT$ to minimum observed mean $RTT$ of any MI in the connection’s history;\cite{nathan-jay2019} and (iii) $plr$ (iv) average throughput of current MI. The next chosen action of the DRL agent is a function of a fixed-length history of the above signal vectors. The reason behind choosing a bounded-length history of signal vectors instead of only the most recent signal vector allows our agent to detect different patterns and
changes in network conditions effectively and react to those changes more appropriately. So, at time $t$ if the state is $s_t$ for our agent, $s_t$ could be defined as:
$s_t=\left(x_{t-(n)}, \ldots, x_{t}\right)$ 
\subsubsection{Actions}
The action space consists of actions that the DRL agent could take to accommodate variations in network conditions. In our design, we only allow five different actions to create a more manageable action space. The first action aggressively increases the number of TCP streams by adding five additional streams; the second action adds only one additional stream; the third action does nothing to the current number of TCP streams; the fourth action reduces the number of TCP streams by removing one stream and finally, the fifth action aggressively reduces the number of streams by a count of five. The choice of these actions is motivated to encourage our DRL agent to quickly increase the number of streams to utilize the available bandwidth while offering an option to decrease the number of TCP streams whenever it is necessary(e.g., congested network). 
\subsubsection{Utility Function and Reward}
We introduce a utility function to specify the overall objective of our DRL agent which is to find an optimal policy that can maximize the cumulative value of the utility function over a transfer session. As discussed in section I, the objective of optimum utilization of available network bandwidth could be decoupled into two parts, maximizing throughput and minimizing packet loss(e.g., avoiding congestion). The choice of minimizing packet loss has two benefits; as an increase in packet loss rate is the most strong signal for network congestion, by minimizing packet loss, we will be able to avoid network congestion. Secondly, incorporating packet loss in our utility function will allow our DRL agent to achieve fairness among other concurrent contesting flows as detecting an event of an increase in packet loss rate will allow our agent to choose an action that can decrease the number of TCP streams.  Our utility function has the following non-linear form:
\begin{equation}
U\left(n_i, T_i, L_i\right)=\frac{T_i}{K^{n_i}}-T_i L_i \times B
\end{equation}
where $n_i$ is the number of parallel TCP streams, $T_i$ is the average throughput and $L_i$ is the aggregate packet loss rate for all TCP streams for $i$th step. $K$ and $B$ are constants that we can tune to control the severity of cost-benefit, performance and punishment terms, respectively. The non-linear utility function guarantees that the transfer will be able to converge and bandwidth resources will be allocated fairly among transfers originating from multiple users, as shown in \cite{arif2021}.

We consider the difference between consecutive utility values from two MIs to devise the reward signal. Instead of the exact value of the utility function, we choose the difference between successive MI's utility value, because the DRL agent's objective dictates that an increase in the utility value points to an improvement and hence the corresponding action should be encouraged. Therefore, we define the reward as follows ~\cite{qtcp}: 
\begin{itemize}
    \item $\mathrm{x}$, if $U_t-U_{t-1}>\epsilon$
    \item $\mathrm{y}$, if $U_t-U_{t-1}<-\epsilon$
\end{itemize}
where $x$ is a positive value and $y$ is a negative value. $x$ and $y$ denote the numerical reward values. $\epsilon$ defines a threshold of changes between utility values. It is a tunable parameter that sets the sensitivity of chosen action for the DRL agent and $\epsilon$ is required for the agent to choose the appropriate action after it finds the current operating optimal point.
\subsection{Training Algorithm}
We use a policy gradient method PPO ~\cite{ppo2017}. PPO employs a clipped surrogate objective and actor-critic style of loss to improve exploration and sample efficiency. 
Algorithm ~\ref{alg:PPO} shows the pseudo-code of training for our agent. The agent runs in the sender node with destination $IP$, $PORT$, and $file directory$ given as input and the objective here is to learn an appropriate policy function $\pi(a|\mathbf{s};\theta)$ for the actor. We first initialize the neural network parameters (i.e., weights and biases) and specify the number of iterations (lines 1-5). It then runs for $N$ episodes to train the value function and policy function respectively(lines 6-21). After each rollout(i.e., episode), the algorithm restarts the file transfer again for a new episode (line 7). As it is on policy algorithm, it follows the current policy to choose an action based on the provided state from the environment and observe the feedback performance signals to calculate rewards. It then adds {$\langle state,action,reward \rangle$} tuple from the current episode to the store object $S$ (line 7-13). After finishing the current rollout, the algorithm resets the gradients (line 14), and then the algorithm iterates over store object $S$ to aggregate all gradients (lines 15-19). After that, the algorithm uses these aggregated gradients to update the parameters of the actor and critic network parameters(i.e., weights and biases) (line 20). It then proceeds to the next rollout (line 21). Line 17 corresponds to policy gradient loss and it defines the direction to update $\theta$ (i.e., actor-network parameters) to increase the overall expected reward. Line 19 shows the critic network's target function $V$ is trained concurrently with the actor network to minimize its estimation error. This critic network gives an estimation of the state value $V(s;\theta_{v})$ which is subtracted from the rollout reward $R$ to reduce the gradient variance. 

\begin{algorithm}[ht]
\SetKwData{Left}{left}
\SetKwData{This}{this}
\SetKwData{Up}{up}
\SetKwFunction{Union}{Union}
\SetKwFunction{FindCompress}{FindCompress}
\SetKwInOut{Input}{Input}\SetKwInOut{Output}{Output}
\SetKwInOut{Subroutines}{Subroutines}
\SetKwComment{comment}{\#}{}
\Input{Receiver $IP$ and $PORT$, file directory $DIR$}
\Output{A stochastic policy function $\pi(a \mid \mathbf{s} ; \theta)$ that outputs a TCP stream selection action $a \in \mathcal{A}$ given a state $s$, and a value function $V\left(\mathrm{~s} ; \theta_v\right)$ that outputs a value estimate for a particular state.}
\BlankLine
\comment{\textbf{Initialization}}
Randomly initialize the model parameters $\theta, \theta_v$\\
Maximum number of episodes $N$\\
\comment{\textbf{Training}}
$n \leftarrow 0$ \\
\While{$n<N$}{
    $s \leftarrow \operatorname{Reset}\left(\mathbf{transferEnvironment}\right)$\\
    $done \leftarrow False$\\
    \While {$done \neq$ True}{
        $s \leftarrow State(transferEnvironment)$\\
        $a \leftarrow  \pi(a|s;\theta)$\\ 
        $S,R \leftarrow \operatorname{Transfer}(s, a)$\\
        \comment{\textbf{added state, action and reward until the episode finishes
        $S=(s, a, R)$ with $s \subseteq Statespace $,$a \subseteq actionspace $ and $r$ is reward. $done$ is $True$ once episode ends}}
        }    
      Reset gradients $d \theta \leftarrow 0$ and $d \theta_v \leftarrow 0$\\
    \For {$(s, a, R) \in$ StateIteratoor $\left(\mathrm{S}\right)$}{
        \comment{\textbf{Accumulate gradients wrt. policy gradient loss}}
        $d \theta \leftarrow d \theta+\nabla_\theta \log \pi(a \mid s ; \theta)\left(R-V\left(s ; \theta_v\right)\right)$\\
        \comment{\textbf{Accumulate gradients wrt. value function loss}} 
        $d \theta_v \leftarrow d \theta _v+\partial\left(R-V\left(s ; \theta_v\right)\right)^2 / \partial \theta_v$
        }
        Perform update of $\theta$ using $d_\theta$ and $\theta_v$ using $d \theta_v$.\\
        $n \leftarrow n+1$
    }
\Subroutines{$\operatorname{Reset}\left(transferEnvironment\right)$: Reset the $transferEnvironment$  to its initial state}
\Subroutines{$\operatorname{State}\left(transferEnvironment\right)$: Get the $transferEnvironment$'s current state}
\Subroutines{$\operatorname{Transfer}(s, a)$: Apply action $a$ to State $s$, and return the
reward achieved}
\Subroutines{$StateIteratoor\left(\mathrm{S}\right)$: Iterate over all the {$\langle state,action,reward \rangle$} tuple stored in $S$ after an episode }

 \caption{Actor-critic(PPO) algorithm for learning a policy to optimize available network bandwidth utilization}
 \label{alg:PPO}

\end{algorithm}
\vspace{-1mm}
We compare the performance of our RL agent with two other online optimization algorithms; (1) Gradient Descent (GD) and (2) Bayesian Optimizer (BO) as used in ~\cite{arif2021}.

\section{Evaluation}
In this section, we will focus on presenting evaluation results for our PPO agent's instantaneous throughput dynamics, throughput performance, fairness metrics, and convergence dynamics of multiple transfers. 
 \subsection{Experimental Setup}
We trained and evaluated our learning-based model in two different wide-area network testbeds. The testbeds are (1) Chameleon Cloud ~\cite{keahey2020lessons}, server located at the University of Chicago and client located at the Texas Advanced Computing Center; (2) CloudLab ~\cite{Duplyakin+:ATC19}, server located at the University of Wisconsin and client located at the University of Utah. Both the Chameleon nodes run on either a Dell PowerEdge R630 containing 24 CPU cores distributed in dual-socket Intel Xeon E5-2670 v3 "Haswell" processors, each containing 12 cores, or on a PowerEdge R740 containing two Intel Xeon Skylake CPUs (each with 12 cores / 24 threads). The client within the CloudLab architecture runs on an HPE ProLiant XL170r server containing 10 CPU cores plus hyper-threading distributed in an Intel E5-2640v4 "Broadwell" processor having 64 GiB of RAM. All the transfers were done using $500 \times 1GB$ files and all these transfers were memory-to-memory types. 
\subsection{Instantaneous Throughput Dynamic of Single Transfer}
To demonstrate the performance in terms of how soon the optimization algorithms could find an optimum point in the search space and how algorithms behave after finding the optimal points, we limited each TCP stream's achievable throughput to 50 Mbps (throttling). Additionally, we limit only sender and receiver nodes to be present in the local network (i.e., no-background traffic). So if the network bandwidth is 1000Mbps, we need twenty parallel TCP streams (i.e., CC) to fully utilize the link. If we create more than twenty streams we would also be able to yield 1000Mbps throughput, but it will result in a lower utility value because of the increment in packet loss rate and the divisor term in equation 3 (i.e.,${K^{n_i}}$). We run all three algorithms each one running for 200 seconds.

\begin{figure*}[ht]
  \begin{subfigure}[b]{0.5\textwidth}
    \includegraphics[width=\columnwidth]{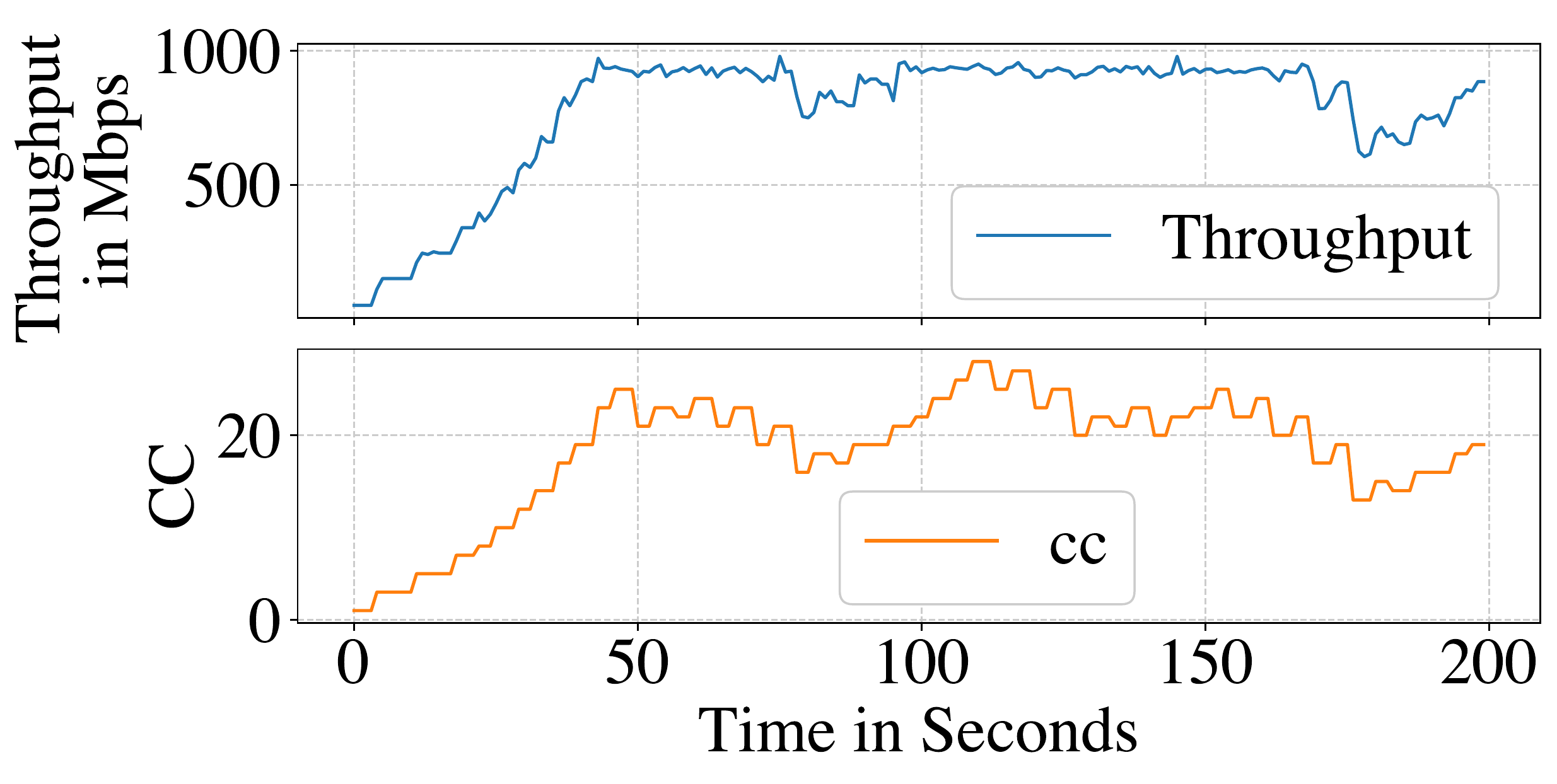}
    \caption{Gradient Descent}
    \label{fig:GD}
  \end{subfigure}
  \hspace{-0.20cm}
  \begin{subfigure}[b]{0.5\textwidth}
    \includegraphics[width=\columnwidth]{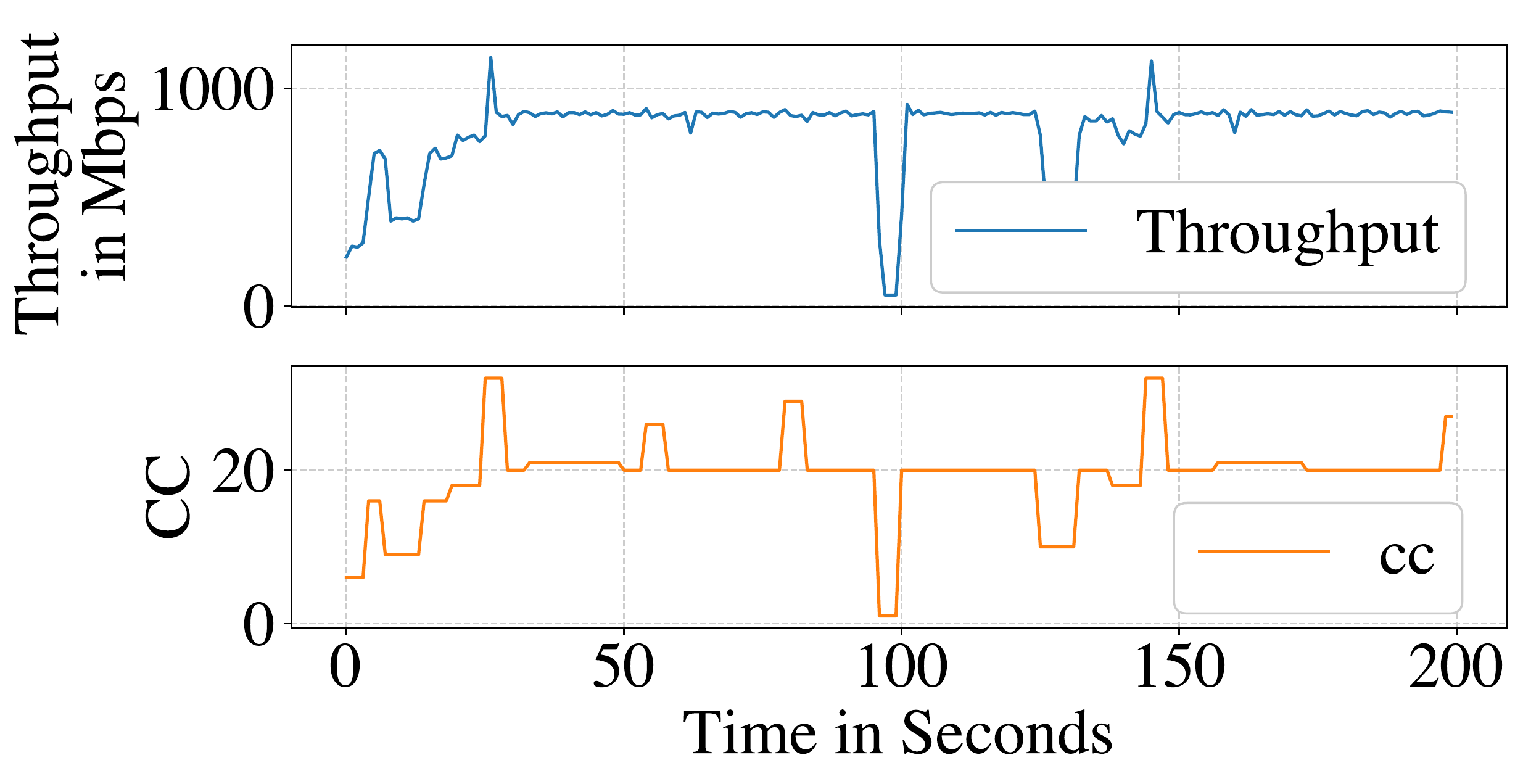}
    \caption{ Bayesian optimizer }
    \label{fig:BO}
  \end{subfigure}
     \hfill 
  \begin{subfigure}[b]{0.5\textwidth}
    \includegraphics[width=\columnwidth]{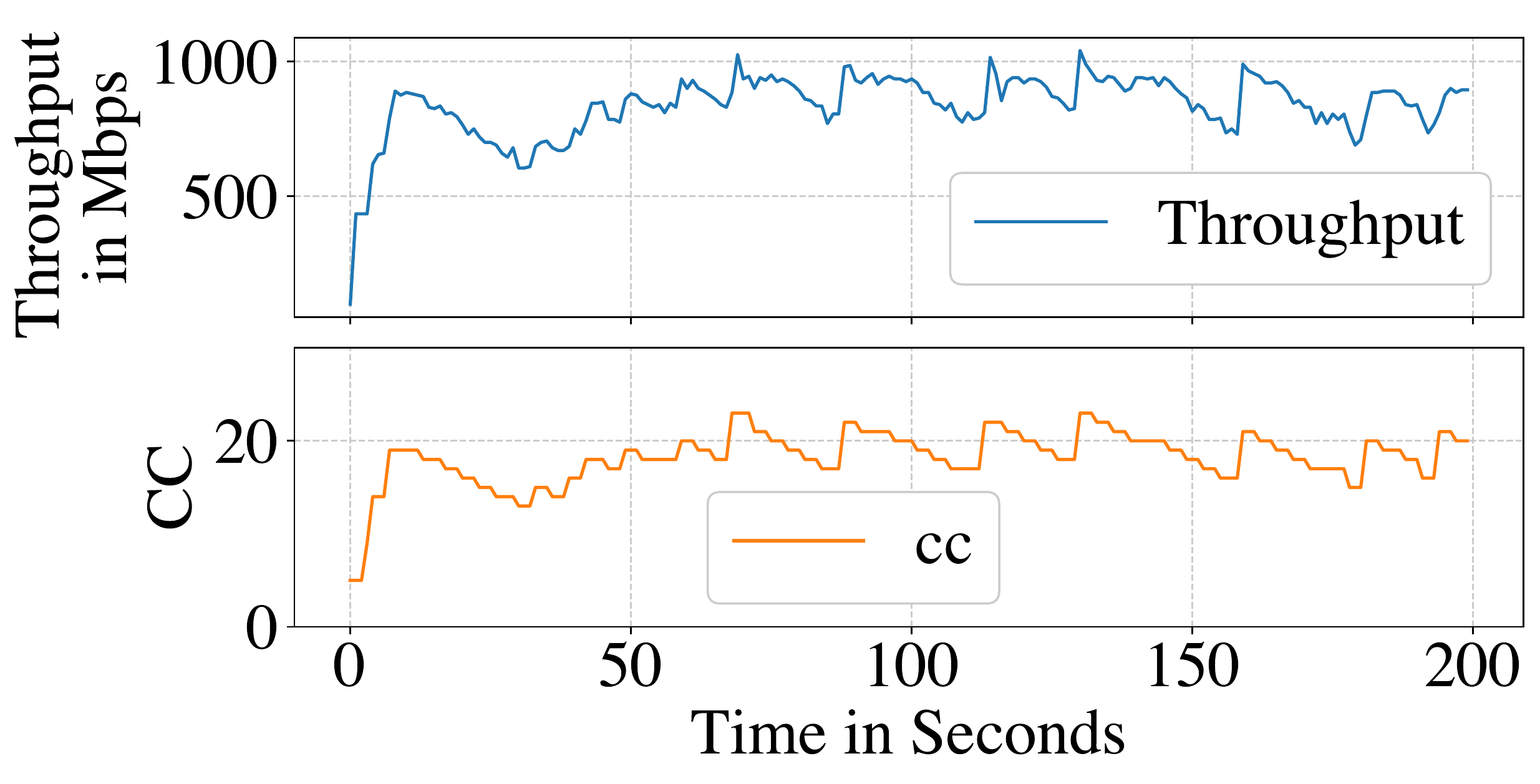}
    \caption{Reinforcement learning (PPO) }
    \label{fig:RL}
  \end{subfigure}
  \hspace{-0.20cm}
  \begin{subfigure}[b]{0.5\textwidth}
    \includegraphics[width=\columnwidth]{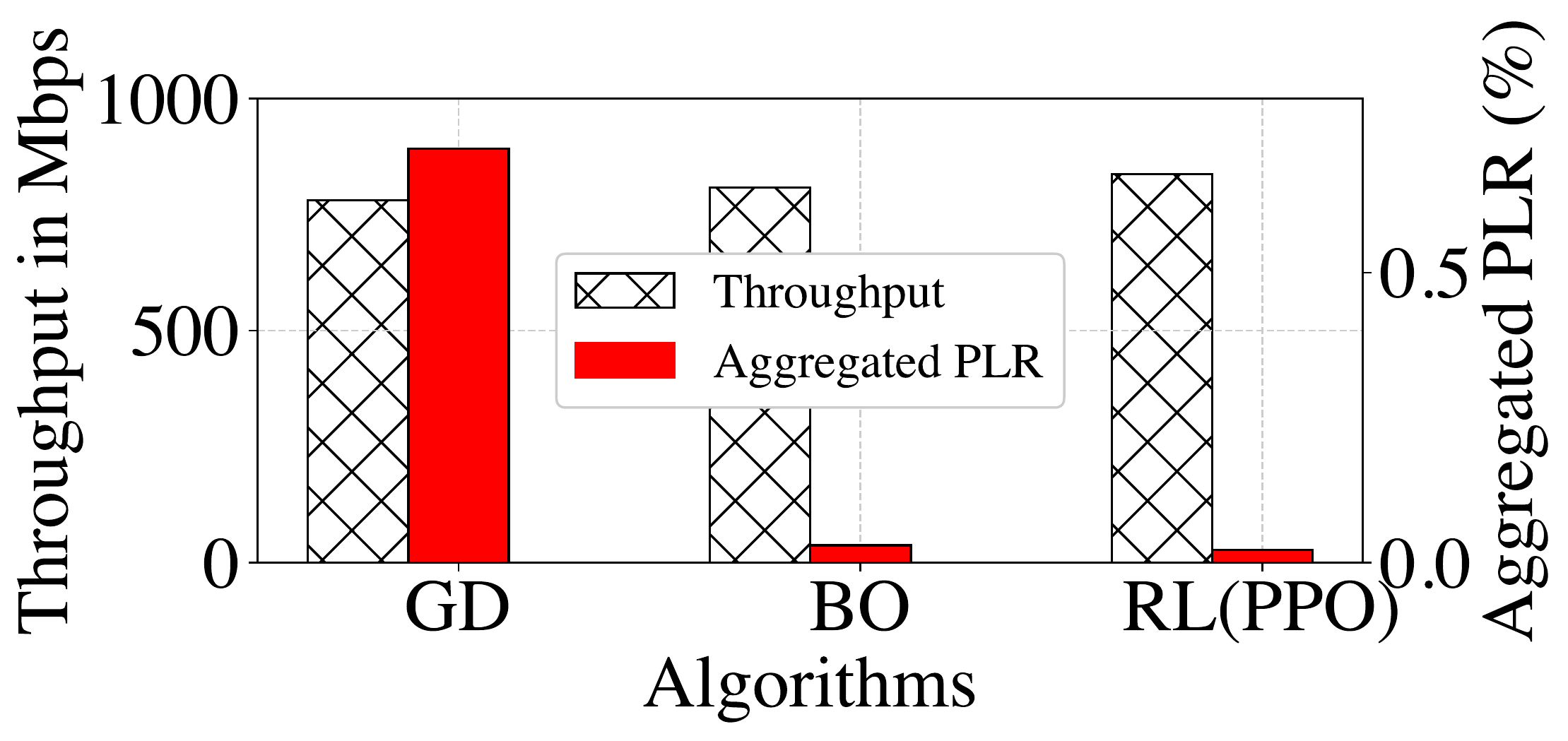}
    \caption{PLR and Throughput}
    \label{fig:GD_BO_RL_PLR_THROUGHPUT}
  \end{subfigure}
  \caption{Performance (i.e., throughput and aggregated packet loss rate) comparison of Gradient Descent, Bayesian Optimization, and Reinforcement Learning (PPO) algorithm where the optimal number of the parallel stream (i.e., CC) is 20}
  \label{fig:figure2}
  
\end{figure*}

As shown in Figure~\ref{fig:GD}, Gradient Descent (GD) achieves peak performance in 50 seconds after starting with CC=2. After finding the optimal point, it keeps checking higher and lower values around the optima to detect any changes in the network condition. Bayesian optimizer (BO) reaches the optimum point in 30 seconds (Figure~\ref{fig:BO}). BO learns the optimal search region after a few initial random sampling and focuses on a concurrency value of 20. Because the surrogate model of BO uses a limited number of past observations, even after converging, BO periodically runs search space exploration. The RL agent was trained and it took 5,000 episodes (i.e., 28 hours of training time with 20-second episodes) to converge. When it is evaluated, the RL (PPO) agent found the optimal operating point in 12 seconds as shown in Figure~\ref{fig:GD}). Upon finding the optimum point, RL agent learns to keep CC value bouncing around 20 so it keeps evaluating a different number of TCP streams to detect changes in the optimal point. Figure~\ref{fig:GD_BO_RL_PLR_THROUGHPUT} shows that, for the  $500 \times 1GB$ file transfer job, RL agent was able to achieve 837Mbps throughput while BO and GD were able to get 807Mbps and 780Mbps respectively. Additionally, RL agent achieves the least amount of packet loss rate aggregated over the transfer duration as shown in Figure~\ref{fig:GD_BO_RL_PLR_THROUGHPUT}. 

\begin{figure}[ht]
\begin{center}
    \begin{subfigure}[b]{0.48\textwidth}
    \includegraphics[width=\columnwidth]{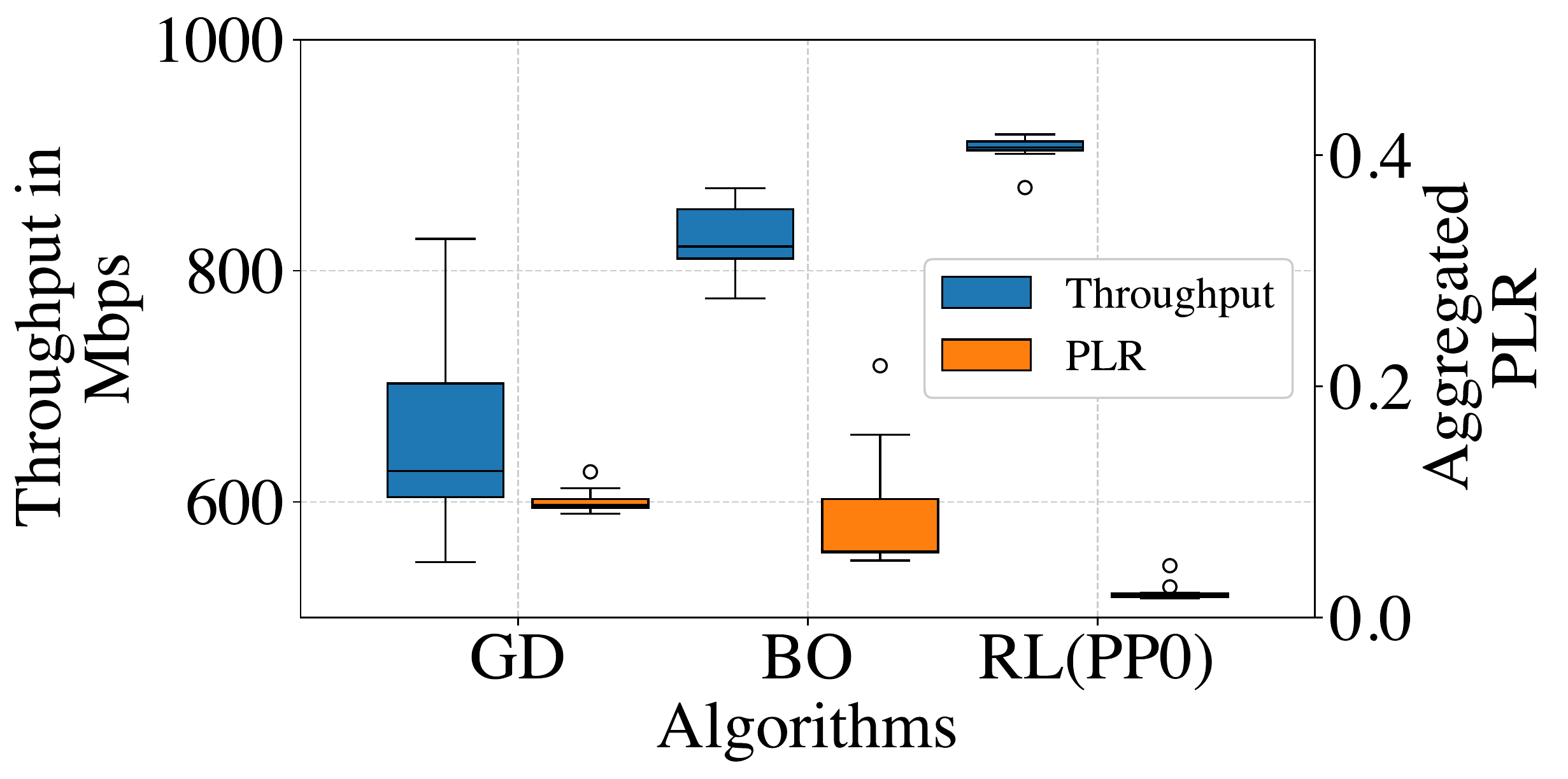}
    \caption{Cloudlab}
    \label{fig:actualTransferCloudlab}
  \end{subfigure}
  \hfill 
     \hfill 
  \begin{subfigure}[b]{0.48\textwidth}
    \includegraphics[width=\columnwidth]{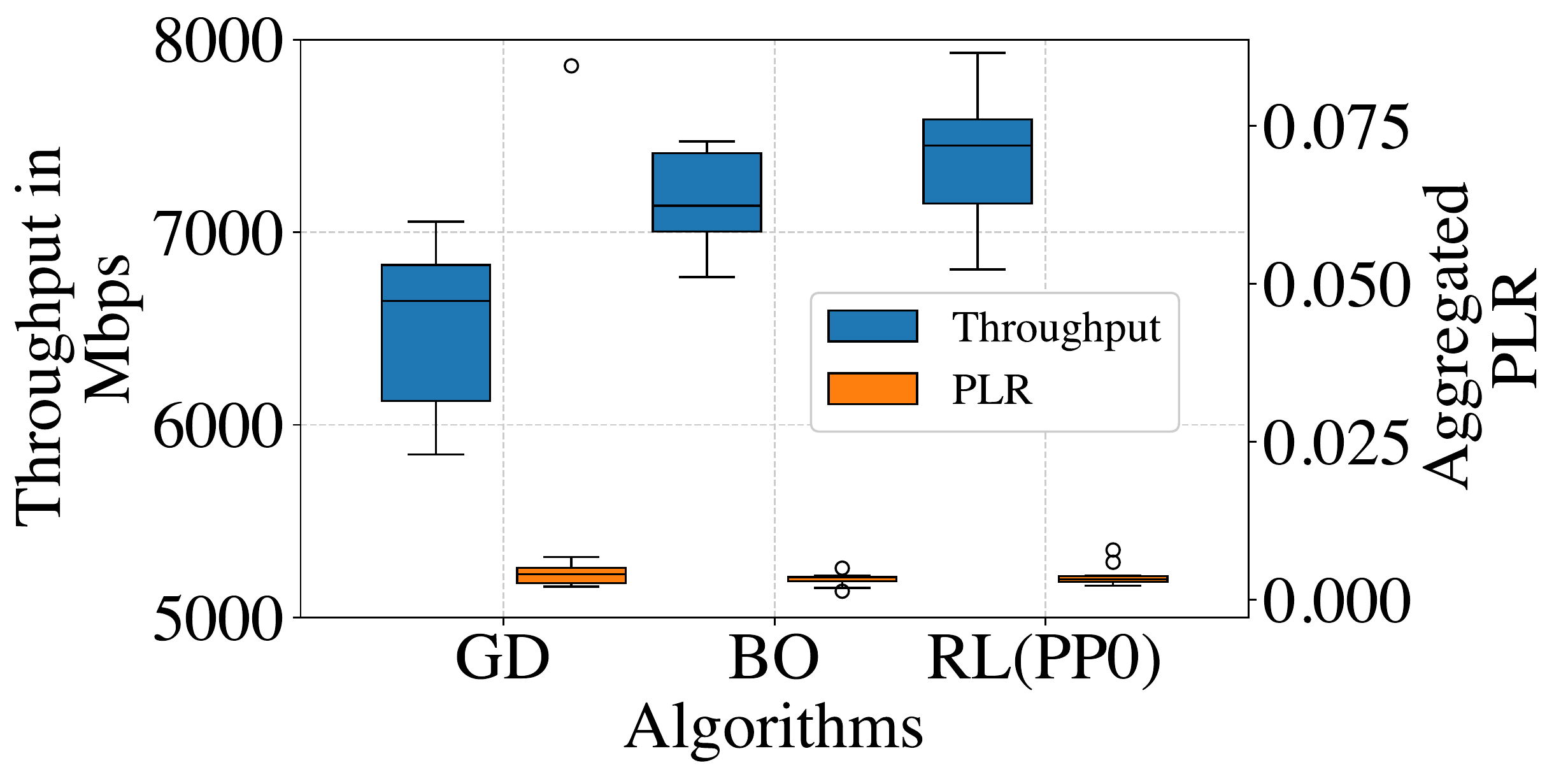}
    \caption{Chameleon}
    \label{fig:actualTransferChameleon}
  \end{subfigure}
  \caption{Boxplot showing throughput and aggregated packet loss rate for different algorithms in different testbeds}
  \label{fig:figure3}
\end{center}
\end{figure}

The above-reported results are expected as RL agent can learn and generalize patterns in network conditions change from the network signal vectors and react to these changes by adjusting the number of CC efficiently. The non-linear form of the utility function allows the RL agent to learn how to avoid actions that lead to network congestion while maximizing the utilization of available network bandwidth. This same RL agent is retrained for 1000 additional episodes for cloudlab and chameleon testbed respectively and we remove the bandwidth limitation(throttling) of TCP streams. The retraining serves the purpose of fine-tuning the RL agent's policy to adapt to the new testbed environment and change in bandwidth capacity (e.g., cloudlab has 1Gbps link capacity whereas Chameleon has 10 Gbps). Figure~\ref{fig:figure3} shows the cloudlab and chameleon testbed performance for 10 runs for each algorithm where each run transfer a total $500 \times 1GB$ files. In both testbeds, as shown in the box plots, RL agent was able to achieve around 15\% more mean throughput while maintaining a lower aggregated packet loss rate.   

\subsection{Fairness and Convergence of Multiple Transfer}
\begin{figure*}[ht]
  \begin{center}
    \begin{subfigure}[b]{0.48\textwidth}
    \includegraphics[width=\columnwidth]{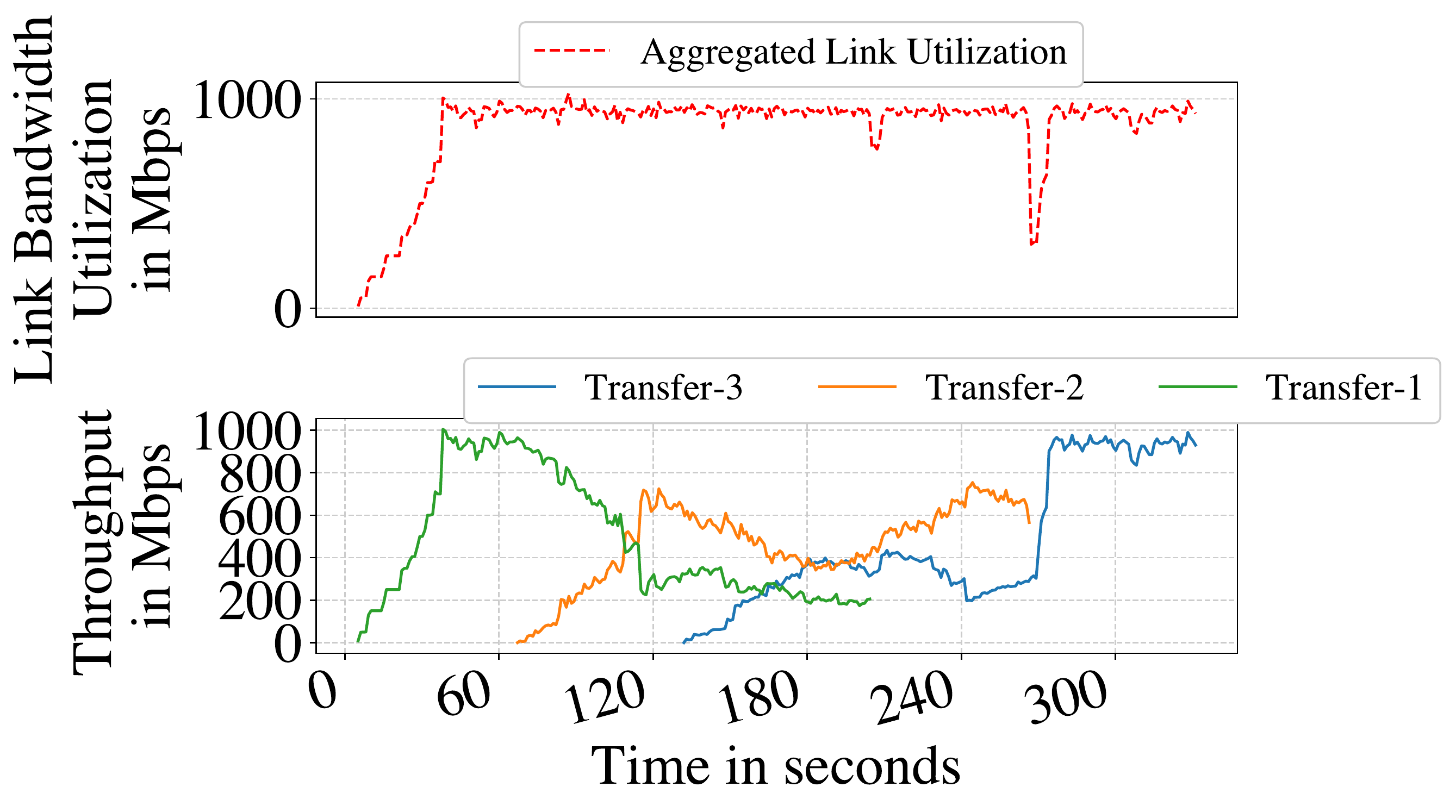}
    \caption{Gradient Descent Optimizer}
    \label{fig:cloudlab_GD}
  \end{subfigure}
  \begin{subfigure}[b]{0.48\textwidth}
    \includegraphics[width=\columnwidth]{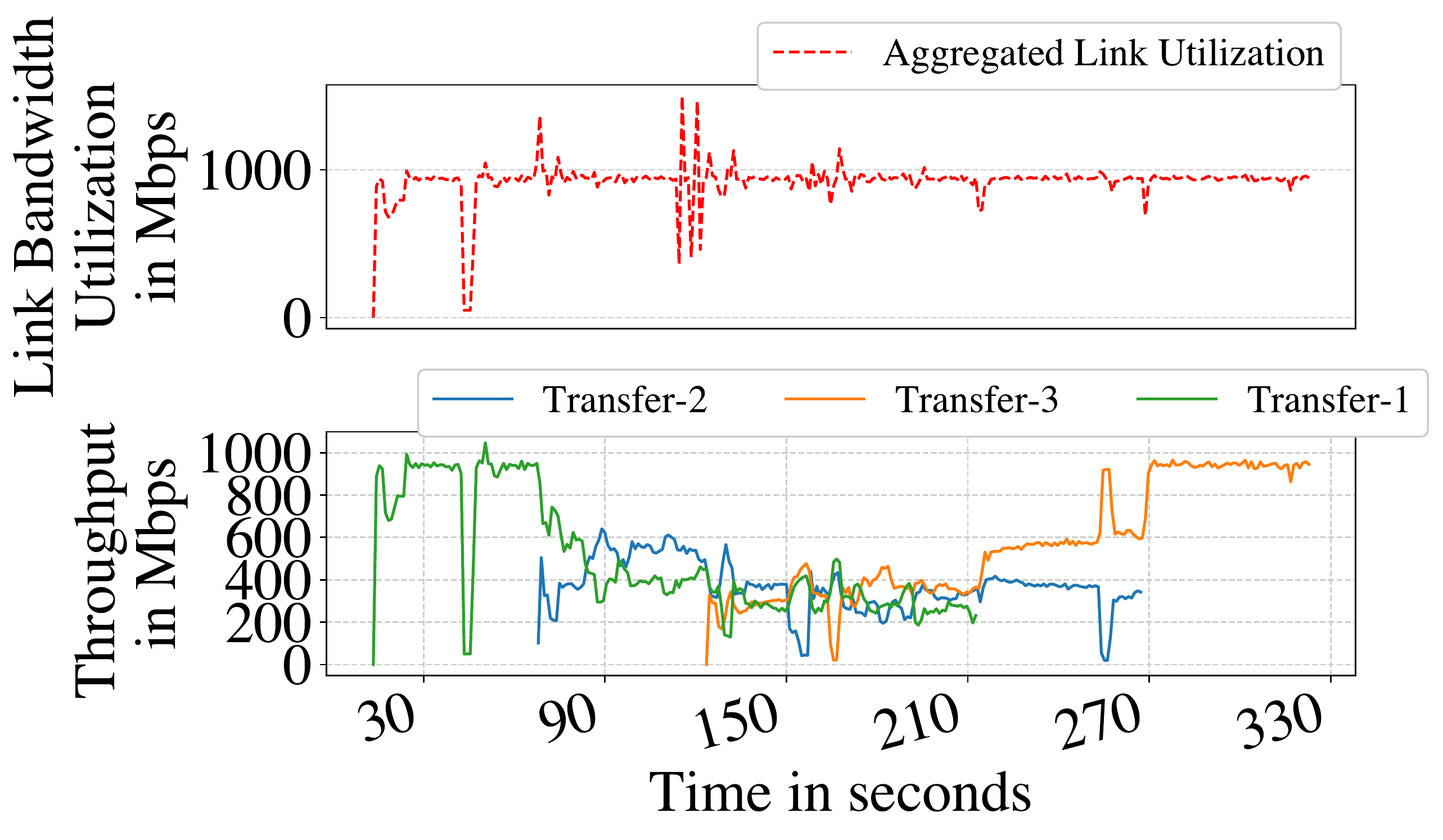}
    \caption{Bayesian Optimizer}
    \label{fig:cloudlab_BO}
  \end{subfigure}
  \begin{subfigure}[b]{0.48\textwidth}
    \includegraphics[width=\columnwidth]{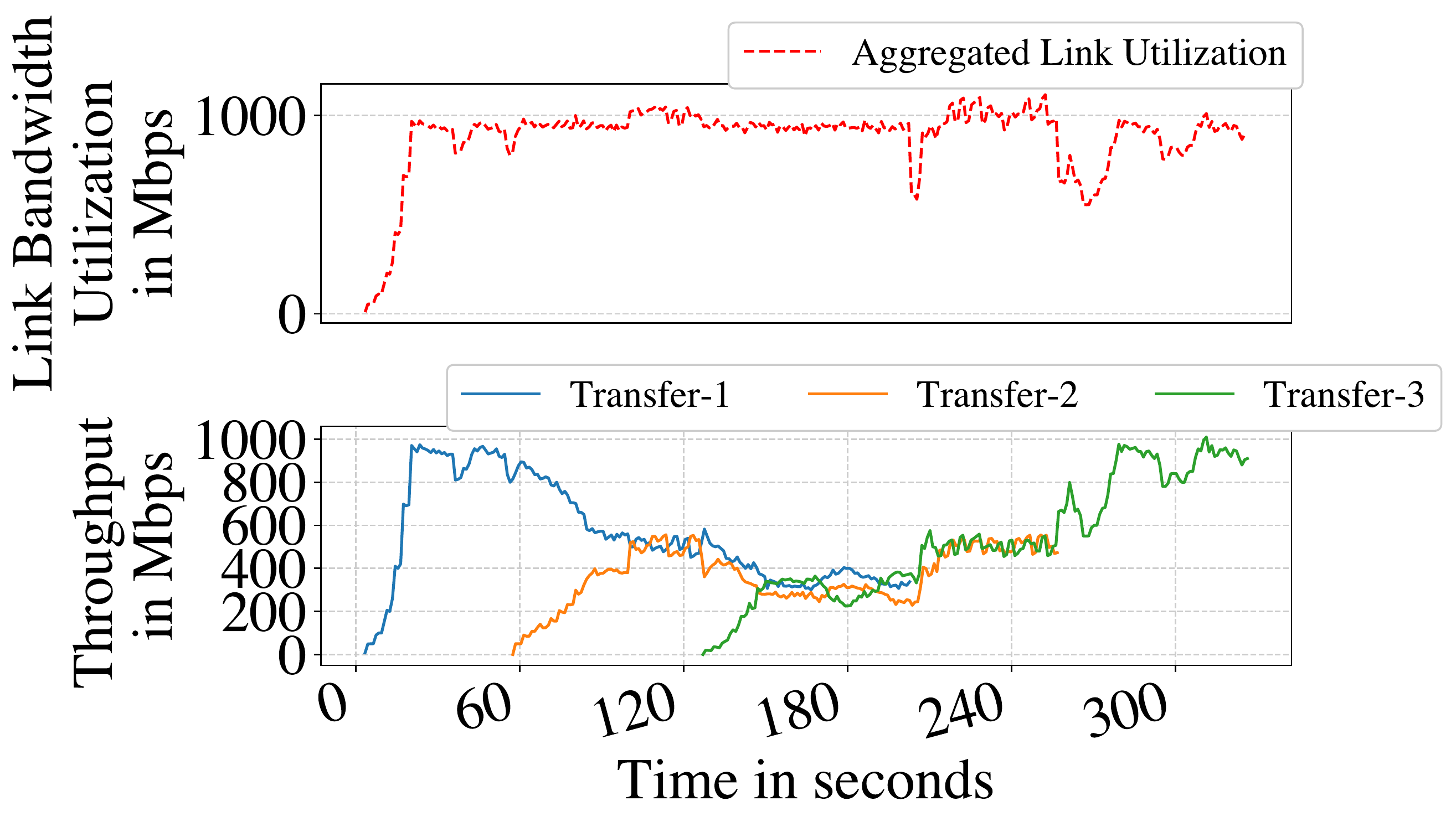}
    \caption{DRL (PPO)}
    \label{fig:cloudlab_RL}
  \end{subfigure}
    \begin{subfigure}[b]{0.48\textwidth}
    \includegraphics[width=\columnwidth]{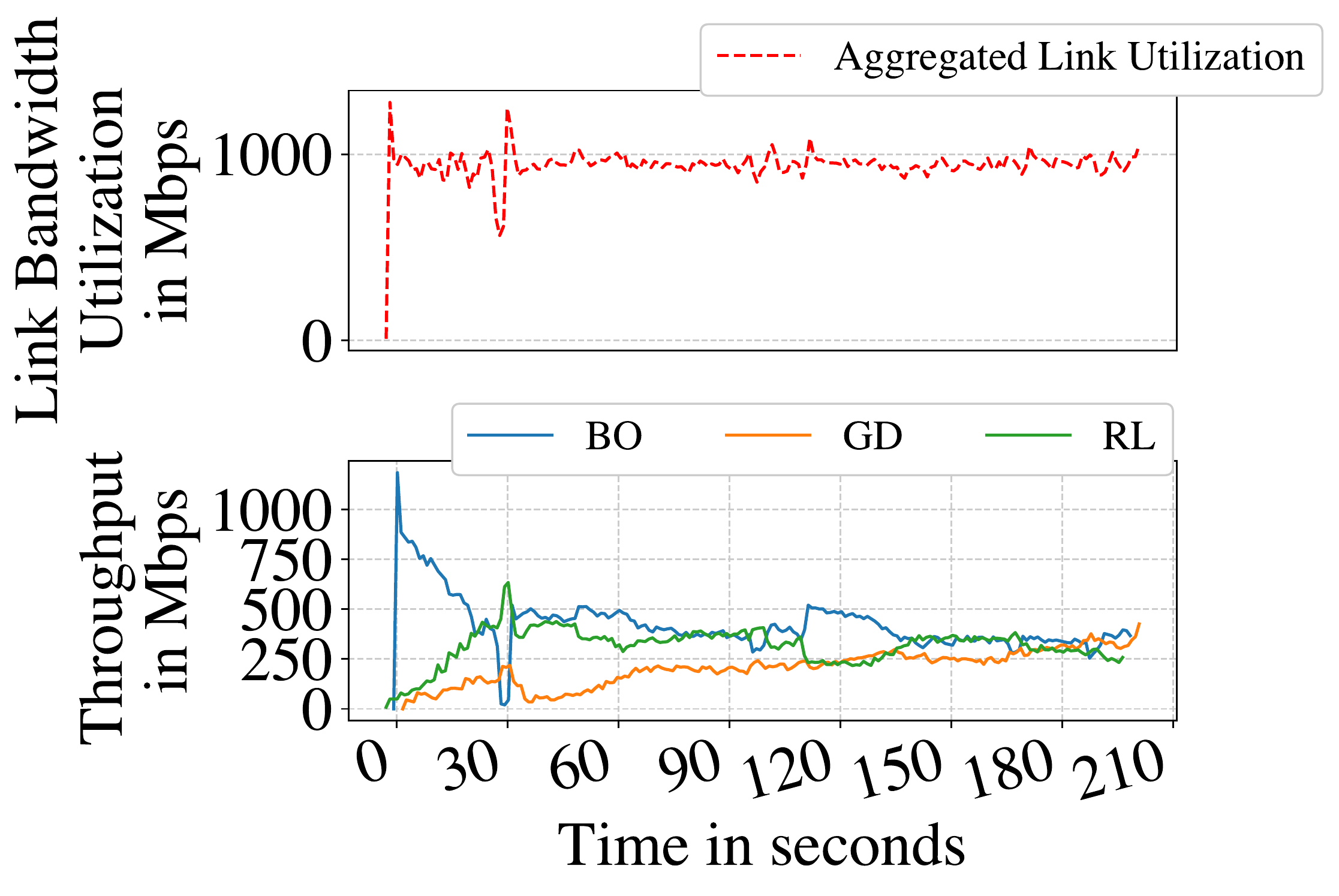}
    \caption{ Concurrent BO, GD, DRL}
    \label{fig:cloudlab_all_concurrent}
  \end{subfigure}
  \caption{ Dynamic Throughput while multiple transfers share same network resource in cloudlab with 1 Gbps total link bandwidth]}
  \label{fig:figure4}
\end{center}
\end{figure*}

Finally, we assess fair resource sharing and convergence dynamics of the three algorithms in cloudlab testbed while multiple transfers were initiated over a single bottleneck link and each transfer is controlled by a separate instance of a different optimizer (i.e., either GD, BO or RL). We also limited each TCP stream’s achievable throughput to 50 Mbps (throttling). As shown in Figure~\ref{fig:cloudlab_GD}, three transfers, all employing GD optimizer were started tentatively 50 seconds apart from the preceding one. When transfer 2 starts, the instantaneous throughput of transfer 1 drops but transfer 1 and 2 never converges. When transfer 3 starts, both transfer 1 and 2 instantaneous throughput drops but their throughput did not converge to a stable value either.  Transfer 1, 2 and 3 achieve 471Mbps, 465Mbps, and 479Mbps average throughput for 200-second run respectively and the average link bandwidth usage is 866Mbps. BO achieves better performance in terms of fair resource sharing and convergence than GD as shown in Figure~\ref{fig:cloudlab_BO}. Transfer 2 can almost reach up to transfer 1's instantaneous throughput level although there are small differences between their throughput as shown after timestamp 90. When transfer 3 joins, both transfer 1 and 2 throughput drops to accommodate transfer 3 but there are occasional fluctuations in different transfers throughput between timestamp 150 and 210. Transfer 1,2 and 3 achieve 492Mbps, 369Mbps, and 566Mbps average throughput for 200-second run respectively and average link bandwidth usage is 921Mbps. On the other hand, when three independent RL agent starts sharing the link, they converge to the fair share of the available link bandwidth as shown in Figure~\ref{fig:cloudlab_RL}. Transfer 1, 2 and 3 achieve 554Mbps, 364Mbps, and 524Mbps average throughput for 200-second run respectively and average link bandwidth usage is 893Mbps. BO achieves higher link bandwidth usage by aggressively putting the network in a congested state as shown in Figure~\ref{fig:cloudlab_BO} at timestamp 150. The reason RL achieves better convergence and a fair share is due to the incorporation of the bounded historical network signal vectors in RL agent's state so RL agent could detect patterns in network changes quickly to adjust the number of TCP streams to avoid network congestion. Finally, in Figure~\ref{fig:cloudlab_all_concurrent}, all three different algorithms, running three different transfers were instantiated concurrently. RL and BO can fairly share the network bandwidth while GD takes significantly longer to converge. As shown around timestamp 180, GD converges and all three transfers can fairly share the available network bandwidth. Transfer using GD optimizer achieves 205Mbps, BO achieves 420Mbps and RL(PPO) algorithm achieves 316Mbps average throughput during this 200 seconds experiment and the average link bandwidth usage is 941Mbps.

\begin{figure*}[ht]
  \begin{center}
    \begin{subfigure}[b]{0.48\textwidth}
    \includegraphics[width=\columnwidth]{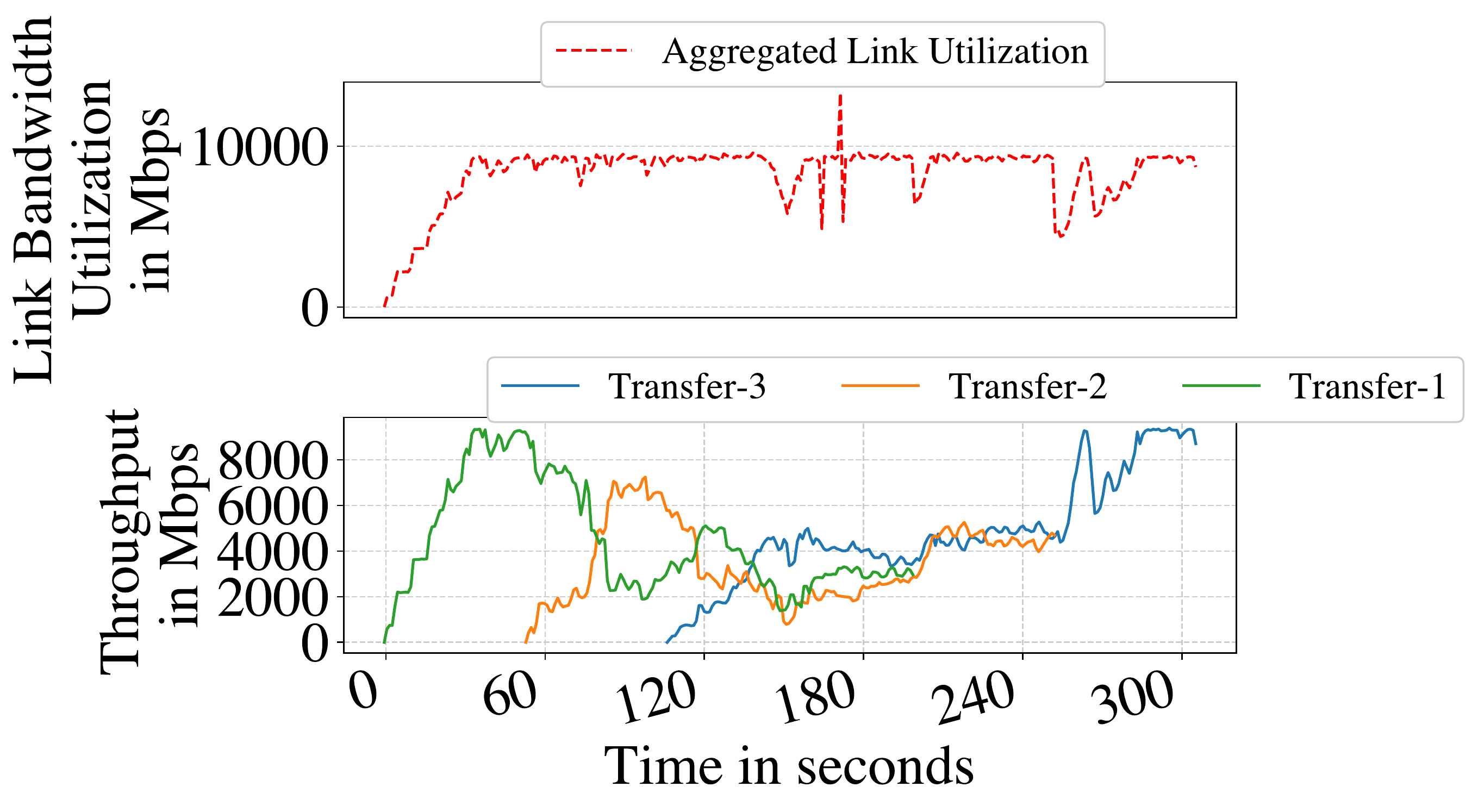}
    \caption{Gradient Descent Optimizer}
    \label{fig:chameleon_GD}
  \end{subfigure}
  \begin{subfigure}[b]{0.48\textwidth}
    \includegraphics[width=\columnwidth]{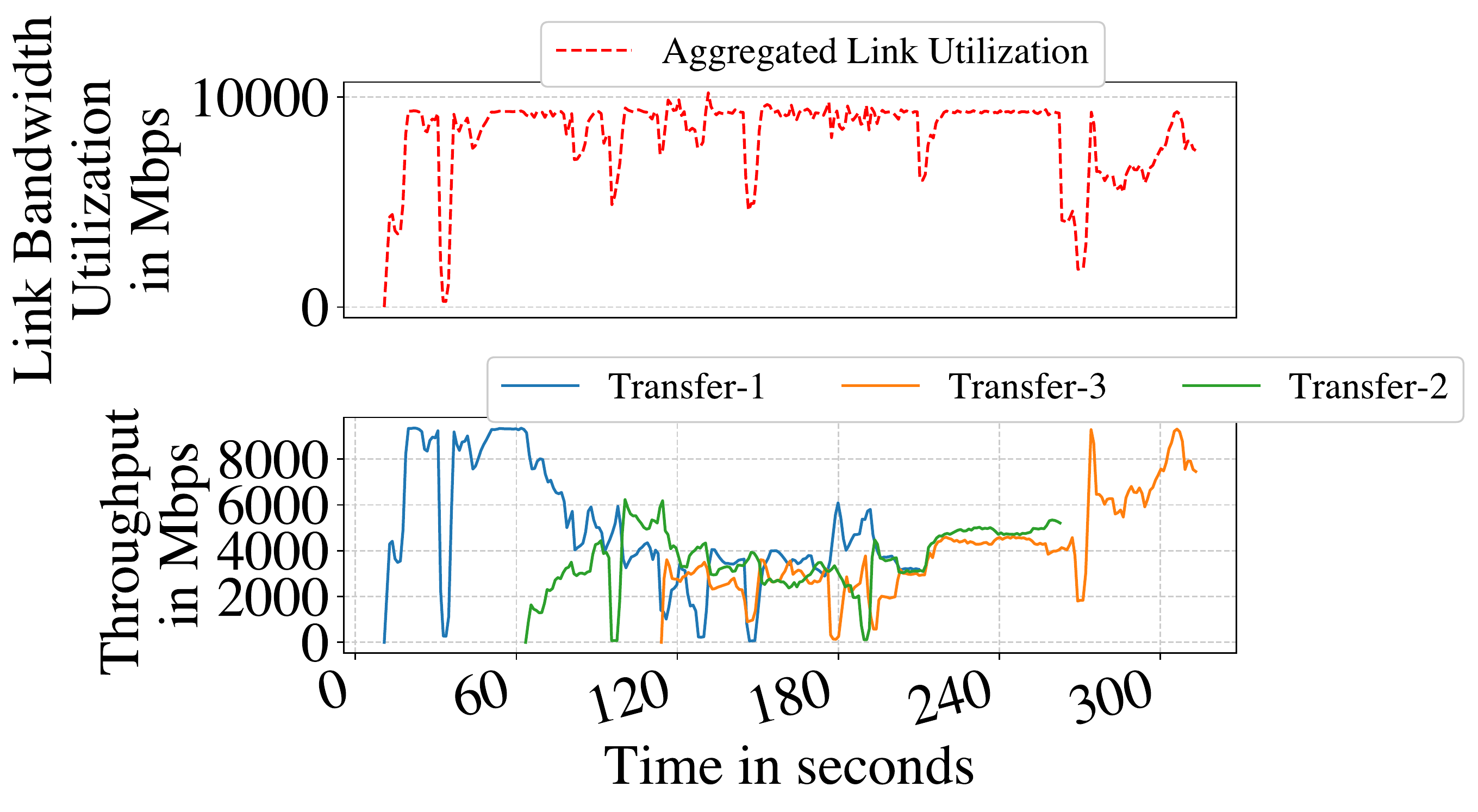}
    \caption{Bayesian Optimizer}
    \label{fig:chameleon_BO}
  \end{subfigure}
  \begin{subfigure}[b]{0.48\textwidth}
    \includegraphics[width=\columnwidth]{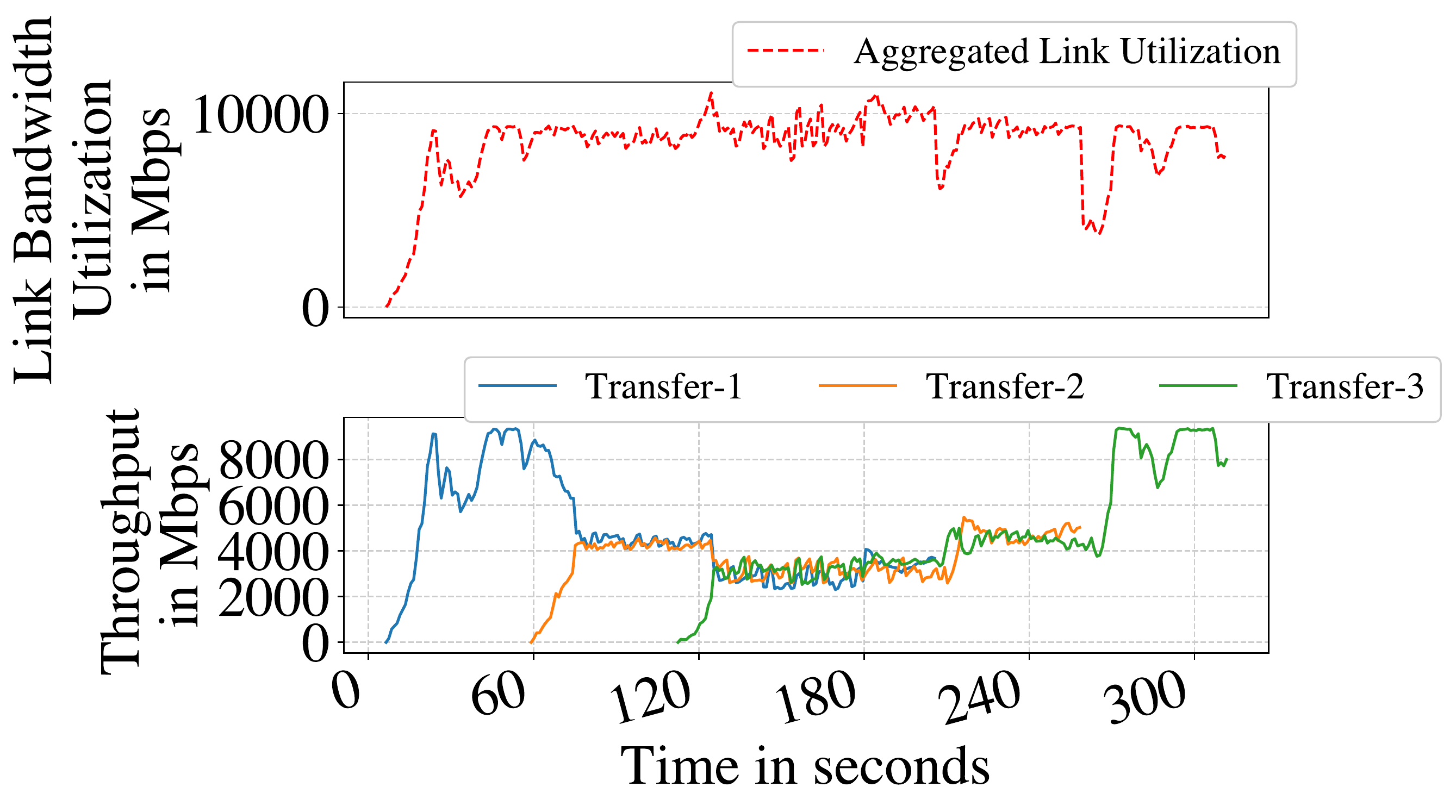}
    \caption{DRL (PPO)}
    \label{fig:chameleon_RL}
  \end{subfigure}
    \begin{subfigure}[b]{0.48\textwidth}
    \includegraphics[width=\columnwidth]{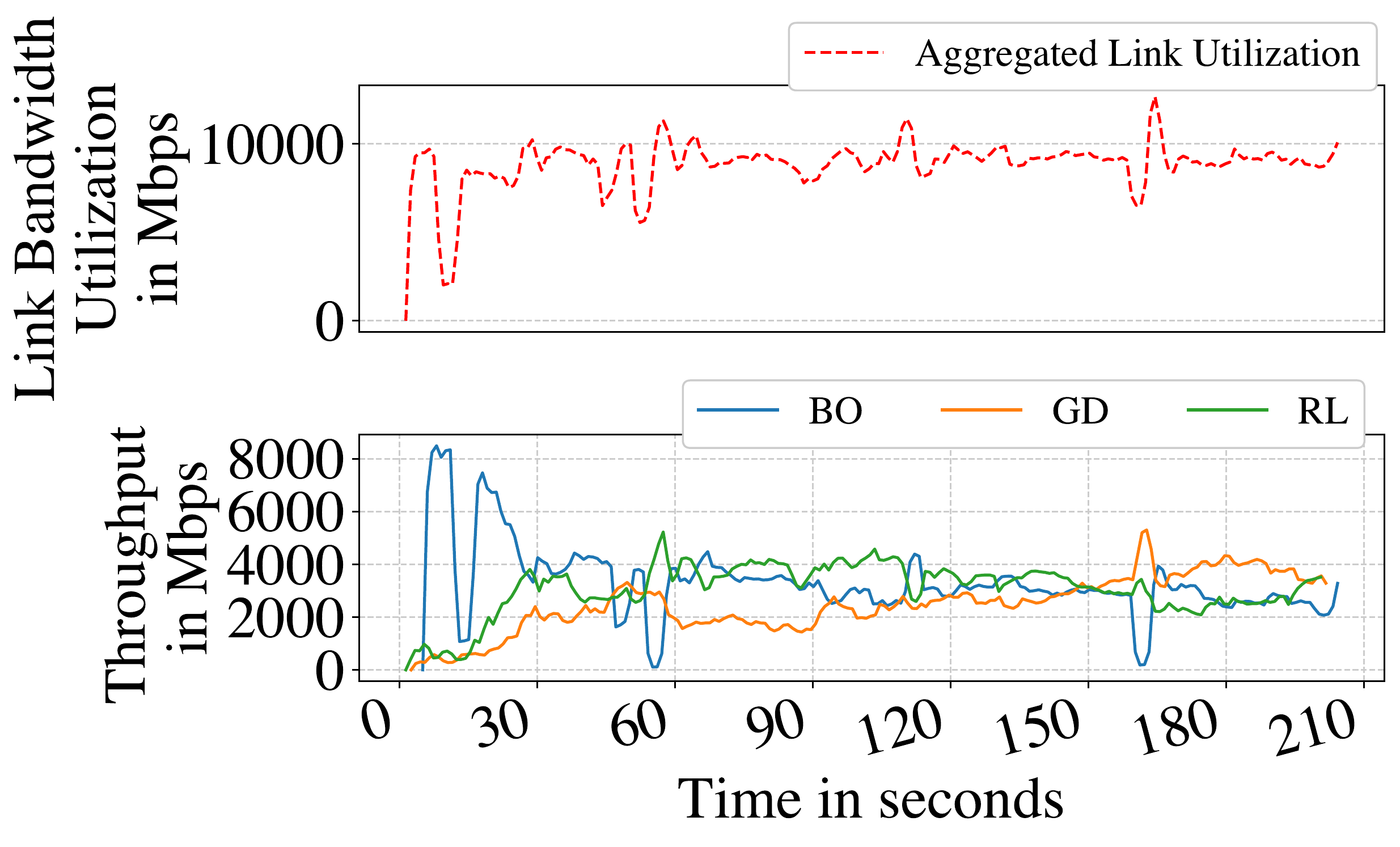}
    \caption{ Concurrent BO, GD, DRL}
    \label{fig:chameleon_all_concurrent}
  \end{subfigure}
  \caption{ Dynamic Throughput while multiple transfers share same network resource in chameleon with 10 Gbps total link bandwidth]}
  \label{fig:figure5}
\end{center}
\end{figure*}

\begin{figure}[ht]
    \centering
    \includegraphics[width=0.48\textwidth]{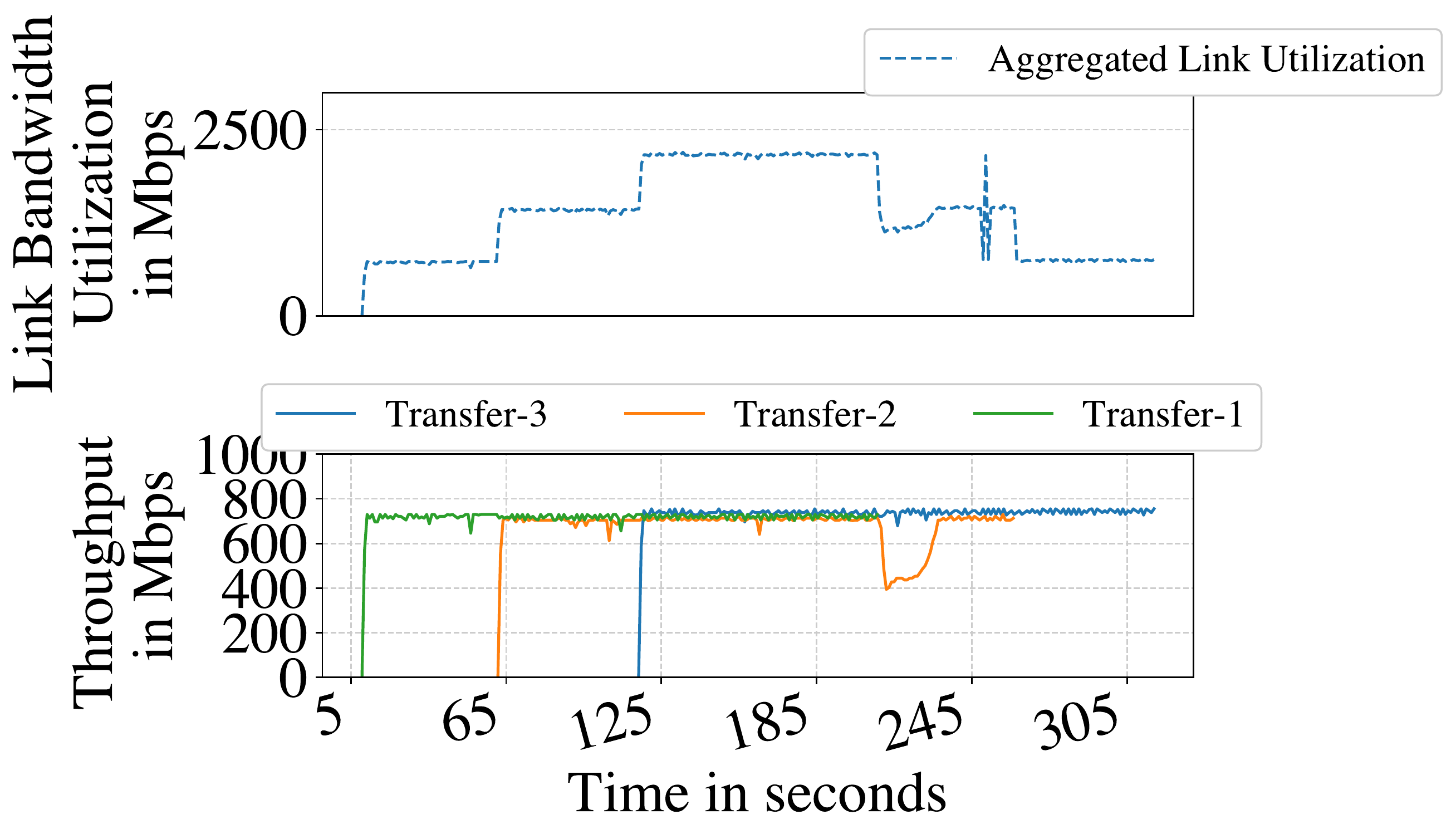}
    \caption{Dynamic Throughput while multiple transfers with NO optimization share same network resource in chameleon with 10 Gbps total link bandwidth}
    \label{fig:figure6}
\end{figure}

Additionally, we assess fair resource sharing and convergence dynamics of the three algorithms in chameleon testbed while multiple transfers were initiated over a single bottleneck link and each transfer is controlled by a separate instance of a different optimizer (i.e., either GD, BO or RL). In these experiments, we removed TCP throughput throttling as the link bandwidth available in chameleon is 10Gbps and multiple TCP streams is required to saturate the link. As in cloudlab, the link bandwidth is 1Gbps, without TCP throughput throttling, the efficacy of different optimizers can not be realized. As shown in Figure~\ref{fig:chameleon_GD}, three transfers, all employing GD optimizer were started tentatively 50 seconds apart from the preceding one. When transfer 2 starts, the instantaneous throughput of transfer 1 drops but transfer 1 and 2 never converges. When transfer 3 starts, both transfer 1 and 2 instantaneous throughput drops but their throughput did not converge to a stable value either.  Transfer 1,2 and 3 achieve 4889Mbps, 3629Mbps, and 3945Mbps average throughput for a 200-second run respectively and average link bandwidth usage is 8200Mbps. For chameleon testbed, BO also achieves slightly better performance in terms of fair resource sharing and convergence than GD as shown in Figure~\ref{fig:chameleon_BO}. Transfer 2 can almost reach up to transfer 1's instantaneous throughput level although there are small differences between their throughput as shown after timestamp 90. When transfer 3 joins, both transfer 1 and 2 throughput drops to accommodate transfer 3 but there are occasional fluctuations in different transfers throughput between timestamp 120 and 180. Transfer 1, 2 and 3 achieve 4456Mbps, 3501Mbps, and 4812Mbps average throughput for 200-second run respectively and the average link bandwidth usage is 8320Mbps. On the other hand, when three independent RL agent starts sharing the link, they converge to the fair share of the available link bandwidth as shown in Figure~\ref{fig:chameleon_RL}. Transfer 1, 2 and 3 achieve 4626Mbps, 3644Mbps, and 4618Mbps average throughput for 200-second run respectively and average link bandwidth usage is 8424Mbps. The reason RL achieves better convergence and fair share \footnote{Note that, as transfer 2 is always sharing the link with other transfers, ideally it should have the lowest average throughput among the three transfers and transfer 1 and 3 should have close to equal throughput as they have the link to themselves for tentatively 50 seconds at the beginning and ending of the transfer respectively.} are due to the incorporation of the bounded historical network signal vectors in RL agent's state so RL agent could detect patterns in the network changes quickly to adjust the number of TCP streams to avoid network congestion. In Figure~\ref{fig:chameleon_all_concurrent}, all three different algorithms, running three different transfers were instantiated concurrently. RL and BO can fairly share the network bandwidth while GD takes significantly longer to converge. As shown around 120 timestamp, GD converges and all three transfers can fairly share the available network bandwidth. Transfer using GD optimizer achieves 2486Mbps, BO achieves 3300Mbps and RL(PPO) algorithm achieves 3057Mbps average throughput during this 200 seconds experiment and the average link bandwidth usage is 8843Mbps.
Figure~\ref{fig:figure6} presents the performance of three similar transfers, but this time we did not use any optimizer. Here each transfer uses only one TCP stream (i.e., TCP CUBIC) for the whole duration of the transfer and we can see that with only one TCP stream, a transfer is able to achieve only about 700 Mbps, a fraction of the available network bandwidth. We can see the available link utilization increases when transfer 2 and 3 starts at timestamps 65 and 125, respectively but never reaches to the maximum available bandwidth as shown in Figure~\ref{fig:figure5}. The results again point to the fact that to use available bandwidth effectively, we need optimizers that are capable of quickly finding and converging to available network bandwidth without congesting the network and also ensuring fairness among other contending transfers and background traffic.  



\section{Conclusion}
Using parallel TCP streams allows an application to use a large virtual message segment size (MSS) to achieve higher throughput. But, using too many TCP streams arbitrarily could create congestion in the network and introduce unfairness. Additionally, the optimum number of parallel TCP streams at a given time is not fixed but dynamic, and the optimum number depends on the nondeterministic background traffic condition. This work shows how we can use DRL to find the optimum number of parallel TCP streams for any transfer. Our learning-based approach is independent of rule-based heuristics and can generalize different network conditions to utilize available network bandwidth intelligently and effectively. After extensively evaluating our RL algorithm with different state-of-the-art online optimization algorithms, we show how our DRL agent could detect patterns in network dynamics from the historical network signal vectors. Our DRL policy-gradient method can find near-optimal points 40\% faster while achieving up to 15\% better throughput than other online optimization algorithms. Furthermore, we also show how incorporating a punishment term in the utility function for the DRL agent's reward derivation, our devised DRL algorithm can avoid network congestion and maintain fairness among other contending transfers.

\section*{Acknowledgements}
This project is in part sponsored by the National Science Foundation (NSF) under award number CCF-2007829.

\bibliographystyle{unsrt}  
\bibliography{references}

\end{document}